\documentclass[aps,pra,groupedaddress,superscriptaddress,notitlepage,twocolumn]{revtex4-1}

\usepackage[english]{babel}
\usepackage[utf8x]{inputenc}
\usepackage[T1]{fontenc}

\usepackage{amsmath,amssymb,amsthm,bm}
\usepackage{bbm}
\usepackage{graphicx}
\usepackage[colorlinks=false]{hyperref}
\usepackage{diagbox}
\usepackage{epigraph}
\usepackage{float}
\usepackage{multirow}
\usepackage{tikz}
\usepackage{mathtools}
\usepackage{braket}
\usepackage{makecell}
\usepackage{array}
\usepackage[all]{xy}
\usepackage[normalem]{ulem}

\newcolumntype{L}{>{\centering\arraybackslash}m{7cm}}
\newcommand{\Rom}[1]{\uppercase\expandafter{\romannumeral#1}}
\newcommand{\rmd}{\mathrm{d}}
\newcommand{\rmi}{\mathrm{i}}

\newcommand{\ex}[1]{\left\langle #1 \right\rangle}
\newcommand{\rme}{\mathrm{e}}

\newcommand{\da}{\dagger}

\newcommand{\pa}{\partial}

\newcommand{\bs}{\boldsymbol}
\newcommand{\mc}{\mathcal}

\DeclareMathOperator{\Tr}{Tr}
\DeclareMathOperator{\sign}{sgn}
\DeclareMathOperator{\Pf}{Pf}
\DeclareMathOperator{\imag}{Im}
\DeclareMathOperator{\real}{Re}

\newcommand{\eqnref}[1]{Eq.\,(\ref{#1})}
\newcommand{\figref}[1]{Fig.\,\ref{#1}}

\definecolor{darkred}{rgb}{0.8,0.1,0.2}

\begin{document}
\title{Magnetic impurities at quantum critical points: large-$N$ expansion and SPT connections}
\author{Shang Liu}
\affiliation{Department of Physics, Harvard University, Cambridge, MA~02138, USA}
\author{Hassan Shapourian}
\affiliation{Microsoft Station Q, Santa Barbara, CA~93106, USA}
\affiliation{Department of Physics, Harvard University, Cambridge, MA~02138, USA}
\affiliation{Department of Physics, Massachusetts Institute of Technology,
 Cambridge, MA~02139, USA}
\author{Ashvin Vishwanath}
\affiliation{Department of Physics, Harvard University, Cambridge, MA~02138, USA}

\author{Max A. Metlitski} 
\affiliation{Department of Physics, Massachusetts Institute of Technology,
 Cambridge, MA~02139, USA}

\begin{abstract}
In symmetry protected topological (SPT) phases, the combination of symmetries and a bulk gap stabilizes protected modes at surfaces or at topological defects. Understanding the fate of these modes at a quantum critical point, when the protecting symmetries are on the verge of being broken, is an outstanding problem. This interplay of topology and criticality must incorporate both the bulk dynamics of critical points, often described by nontrivial conformal field theories, and SPT physics. Here, we study the simplest nontrivial setting - that of a 0+1 dimensional topological mode - a quantum spin - coupled to a 2+1D critical bulk. Using the large-$N$ technique we solve a series of models which, as a consequence of topology, demonstrate intermediate coupling fixed points. We compare our results to previous numerical simulations and find good agreement. We also point out  intriguing connections to generalized Kondo problems and Sachdev-Ye-Kitaev (SYK) models. In particular, we show that a Luttinger theorem derived for the complex SYK models, that relates the charge density to particle-hole asymmetry, also holds in our setting. These results should help stimulate further analytical study of the interplay between SPT physics and quantum criticality. 
\end{abstract}
\maketitle
\tableofcontents
\section{Introduction}

Following the classification of  topological phases of non-interacting fermions in various symmetry settings and dimensions \cite{Kitaev_2009, Ryu_2010,Hasan10, QiZhangRMP}, which include topological insulators and superconductors, attention has turned to  symmetry protected topological (SPT) states in interacting systems. Despite the presence of potentially strong interactions, a key simplification of SPTs is the existence of an energy gap and a unique ground state in the absence of boundaries. There too, significant progress has been achieved \cite{turner2013topological,WenRMP,Senthil_2015} in classifying and characterizing fundamental properties of these phases. SPT phases exhibit many interesting topological phenomena such as robust gapless modes on the surface or on topological defects.

In recent years, the question of stability of topological phenomena to the {\em closing} of the energy gap has begun to be explored \cite{Kestner11,Scaffidi_2017,Kainaris17,Chen18,Verresen18,Jiang18,Parker18,Keselman18,Jones19,verresen_gapless_2019,verresen2020topology}. Surprisingly, various situations where SPT physics survives the closing of an energy gap have been identified, which have been dubbed  gapless SPTs. In particular, examples have been found where the surface modes of an SPT (partially) survive right at the bulk ordering transition, where one of the protecting symmetries is on the verge of breaking spontaneously. This has been shown to lead to unconventional boundary criticality  \cite{Scaffidi_2017,Parker18,Verresen17,Verresen18,verresen_gapless_2019,verresen2020topology}. Most analytical approaches so far focus on 1+1 dimensions where we have greater analytical control of 2D conformal field theories (CFTs) and where numerical simulations are readily accessible. Much less is known in higher dimensions. However, one numerical study of a 2+1D strongly interacting gapless SPT is found in \cite{ZhangFa17} (see also  \cite{DingZhangGuo,Wessel18,WesselSpin1, Guo2021}), where the transition between a 2+1D Affleck-Lieb-Kennedy-Tasaki (AKLT) phase, a topological paramagnet, and a symmetry-breaking phase was studied. Although the bulk criticality was identical to the ordering transition out of a trivial paramagnet, i.e. the $O(3)$ Wilson-Fisher critical point, boundary critical behavior distinct from the ``ordinary" universality class\cite{cardybook, Diehl_1997} of the classical $O(3)$ model was observed. It, however, remains to be understood to what extend this ``non-ordinary" boundary criticality is a consequence of the bulk AKLT physics.\cite{WesselSpin1, Max3DBCFT}

Here, we seek to analyze examples where a gapless boundary or a zero-mode associated with a defect leaves behind a nontrivial fingerprint even after coupling to a critical bulk. In particular, we focus on 2+1D interacting bosonic systems. One may attempt to analyze the setup in the aforementioned numerical work \cite{ZhangFa17,DingZhangGuo,Wessel18,WesselSpin1}: coupling a 1+1D gapless boundary theory originating from the SPT edge modes to the $O(3)$ critical bulk. However, this is a hard problem on two accounts. First, even the problem of boundary critical phenomena in 3D classical statistical mechanics is only now being understood  in models with continuous symmetries \cite{Max3DBCFT, ToldinO3, DengO2}, even in the absence of  SPT physics. Second, one must add quantum effects i.e. SPT physics to the problem. Here we will consider a simpler version of the problem which is both tractable and displays remarkably rich behavior. Instead of terminating the bulk on a 1+1D boundary, we consider a 0+1D topological mode bound to a topological defect in the bulk. Our analysis sheds light on the  closely related  question of the screening of impurity spins at magnetic quantum critical points, and allows us to compare our results with numerical studies on spin models.  

\begin{figure}
    \centering
    \includegraphics{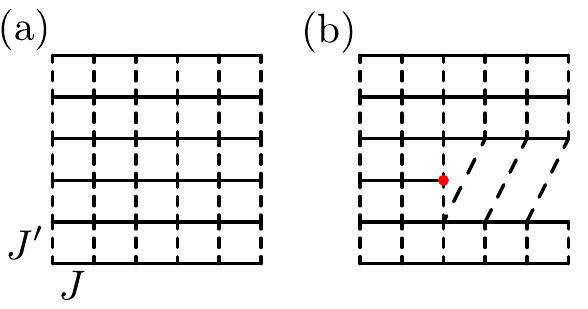}
    \caption{(a) A spin-$1$ antiferromagnetic Heisenberg model. The Heisenberg couplings on horizontal bonds are $J$, while the couplings on vertical bonds are $J'$. When $J'/J$ is sufficiently small, the system is in a topological paramagnetic phase protected by the $SO(3)$ spin rotation symmetry and  translation symmetry. As we increase $J'/J$ up to $1$, there is a transition to the N\'{e}el ordered phase. 
    (b) A dislocation defect with a $(1,1)$ Burgers vector, which supports a spin-$1/2$ topological mode in the paramagnetic phase. }
    \label{DislocationIllustration}
\end{figure}

It is helpful to have a concrete lattice model in mind. Imagine a 2+1D spin-$1$ Heisenberg model $H=\sum_{ij}J_{ij}{\bs S}^i\cdot{\bs S}^j$ with antiferromagnetic couplings as shown in Fig.~\ref{DislocationIllustration}a. The couplings on horizontal bonds are $J$, while the couplings on vertical bonds are $J'$. There are no couplings beyond the nearest-neighbor ones. Now suppose $J'$ is much smaller than $J$, then the system is in a 2+1D AKLT or Haldane phase, a topological paramagnet protected by the $SO(3)$ spin rotation and vertical lattice translation symmetries. Indeed, if we create a vertical edge, then this edge will host a dangling spin-$1/2$ chain which can not be gapped out by adjusting the boundary couplings as long as the spin rotation and vertical translation symmetries are preserved. We may also choose the protecting symmetries to be  spin rotation together with the translation generated by $T_x T_y$ which moves a lattice site to its next-nearest neighbor in the upper right. For the same reason, an edge parallel to the vector $(1,1)$ (compatible with $T_x T_y$) will be gapless given these symmetries. For our purpose here, consider a dislocation defect with a $(1,1)$ Burgers vector as shown in Fig.~\ref{DislocationIllustration}b, which may be regarded as a flux of the translation symmetry generated by $T_xT_y$. It is not hard to see that this defect supports a single dangling spin-$1/2$, a 0+1D topological mode. We can gradually increase the value of $J'$ up to $J$, and at some point there will be a transition from the topological paramagnet phase to the N\'{e}el ordered phase. This transition is in the $O(3)$ universality class as confirmed numerically in Ref.~\onlinecite{Matsumoto2001}. Right at the critical point, the spin-$1/2$ on the defect and the gapless bulk degrees of freedom are strongly coupled to one another, and we inquire about their fate in the low energy limit. This is equivalent to an impurity problem where a spin-$1/2$ impurity is immersed into the 2+1D critical bulk. Two remarks are in order. (i) In a physical lattice, the dislocation will carry a logarithmic elastic energy and will locally distort the couplings far away from the defect core. Here, we will consider an idealized model where locally the couplings away from the defect core are not distorted and one can detect the dislocation only by encircling it. (ii) We have chosen a special dislocation Burgers vector such that the N\'{e}el order parameter $\vec\phi$ acquires no phase by winding around the defect point. In other words, the $O(3)$ critical bulk sees no branch cut due to the dislocation. If the Burgers vector is $(0,1)$, then the dislocation leads to a $\mathbb{Z}_2$ defect line where the order parameter $\vec{\phi}$ twists $\vec{\phi} \mapsto - \vec{\phi}$ across the branch cut. 

We expect the impurity not to be fully screened since the $SO(3)$ bulk symmetry fractionalizes into $Spin(3)=SU(2)$ on the impurity site and the bulk does not have the right degrees of freedom to cancel the spinor impurity. It is then interesting to understand the  dynamics of this impurity, for example its nontrivial spin susceptibility due to interaction with the bulk degrees of freedom. Moreover, one may ask what happens if we replace the spin-$\frac{1}{2}$ impurity by an integer spin eg. spin-$1$; is there an even-odd effect? In this paper, we address these questions using the large-$N$ technique and at the same time present several models with an unscreened impurity which are analogous to the gapless SPT states studied earlier. Our work provides a novel perspective on the interplay between topology and 2+1D quantum criticality. 

The impurity problems of interest in this paper resemble the Kondo effect, especially the multichannel generalizations \cite{Kivelson2CK,AffleckKondoLecture,SUNKondo}. However, an important difference is that the noninteracting bulk electrons are here replaced by strongly coupled critical modes. This further implies that the gapless modes are intrinsically 2+1D and do not admit a dimensional reduction. 

The problem of a spin-$S$ impurity coupled to the $O(3)$ Wilson-Fisher bulk CFT was  analyzed in groundbreaking work in \cite{SachdevScience,SubirImp}, in the context of an impurity spin coupled to critical fluctuations. Utilizing a $D=4-\epsilon$ spacetime dimension expansion, an interacting fixed point near the decoupled one was found for small $\epsilon$, where the impurity is not fully screened and exhibits a Curie form static susceptibility at finite temperature. See also \cite{SubirImpAlternative4DExp,SubirImp2DExp} for alternative approaches. 

In this work, we analyze three $SO(N)$ symmetric generalizations of the above $SO(3)$ problem in the large-$N$ limit. Nontrivial fixed points in which the impurity is not fully screened are found in all these models for the impurity transforming both in the spinor (projective) and non-projective representations of $SO(N)$. To make a contrast, we also consider a fourth slightly different model where a fully-screened fixed point {\em can} be realized for impurities transforming in non-projective representations of $SO(N)$. This is in accord with the expectation that while an impurity transforming projectively under $SO(N)$ can never be screened by the $O(N)$ Wilson-Fisher critical bulk, whether a non-projective impurity is screened depends on the details of the Hamiltonian near the impurity. 

The large-$N$ approach adopted here has the general advantage of generating one-parameter families of new models, being exactly solvable in a strongly-coupled limit and being able to access non-perturbative fixed points (if any). In fact, an $SU(N)$ large-$N$ analysis was also applied in \cite{SubirImp} albeit with a noninteracting bulk. We point out however that in our calculation which includes bulk interactions, a logarithmic UV divergence which is present in the free case is cancelled, leading to finite results for the spin susceptibility. A few other impurity problems with interacting critical bulk were investigated in \cite{FlorensPRL,FlorenPRB}, where the impurity spin is not protected and a fully-screened phase appears in the phase diagram. One impurity model to be studied below is similar to that in \cite{FlorensPRL}, except we study a model with $SO(N)$ rather than $SU(N)$ symmetry.

Finally, let us highlight an unexpected connection. Two of our impurity problems have interesting links to the real and complex Sachdev-Ye-Kitaev (SYK) models \cite{Kit.KITP.1,Kit.KITP.2,SYKRemarks,SubirCplxSYK,GuCplxSYK}, respectively. In particular, in the latter case, we prove a Luttinger theorem relating the spin representation, expressed in terms of a parton `charge density' in the UV to the particle-hole asymmetry of the IR Green's function, which is exactly the same as the so-called charge formula in the complex SYK model \cite{GeorgesParcolletSachdev,SubirCplxSYK,GuCplxSYK}. This is a nontrivial matching because the charge formula contains an anomalous term that is not present, for example, in the multichannel $SU(N)$ Kondo model \cite{SUNKondo}.  

In the next section, we give more detailed motivations and explanations for our impurity models and summarize our main results. The following four sections examine the dynamics of these models in detail. In Section~\ref{ComparisonSection}, we compare with related results including different approaches to the $SO(3)$ problem mentioned above, as well as the results of large scale quantum Monte Carlo numerics,  and the SYK models. We conclude and discuss possible future directions in Section~\ref{Conclusions}. 

\section{Impurity Models and Summary of Results}\label{SO3Problem}
\begin{table*}
	\centering
	\begin{tabular}{|>\centering{c}|>\centering{c}|>\centering{c}|>\centering{c}|L|}
		\hline
		{\bf Sec.} & {\bf Impurity} & {\bf Bulk}
		& {\bf Coupling at} $\boldsymbol{r=0}$ &  \begin{centering}{\bf Low-Energy Properties}\end{centering} \\
		\hline
		
		% Spinor impurity and vector bulk. 
		\ref{SpinorImpVecBulk} & Spinor ($\gamma_0$, $\gamma_{\alpha\geq 1}$) & $\phi_\alpha$ & $(\rmi\gamma_0\gamma_\alpha)\phi_\alpha $ &
        \begin{minipage}{0.4\textwidth}
        \vspace*{10pt}
		\begin{flushleft} 
			$[\gamma_0]=1/2$, $[\gamma_{\alpha\geq1}]=0$, $\chi_{\rm imp}={\mc C}_{\rm cp}/T$ (cf. Eq.\,\ref{chiResult_SpinorImpVecBulk}). \\
			
			Impurity not fully screened. 
		\end{flushleft}
        \vspace*{3pt}
        \end{minipage}
        \\
		\hline
		
		% Spinor impurity and tensor bulk. 
		\ref{SpinorImpTensorBulk} & Spinor ($\gamma_0$, $\gamma_{\alpha\geq 1}$) & $\phi_{\alpha\beta}$ & $(\rmi\gamma_\alpha\gamma_\beta)\phi_{\alpha\beta} $ & 
        \begin{minipage}{0.4\textwidth}
        \vspace*{10pt}
		\begin{flushleft} 
			$[\gamma_0]=0$, $[\gamma_{\alpha\geq1}]=1/4$, $\chi_{\rm imp}={\mc C}_{\rm cp}/T$ (cf. Eq.\,\ref{chiResult_SpinorImpTensorBulk}). \\
			Impurity not fully screened. 
		\end{flushleft}
        \vspace*{3pt}
        \end{minipage}
        \\
		\hline
		
		% Tensor impurity and tensor bulk. 
		\ref{TensorImpTensorBulk} & \makecell{Antisym. $\nu N$-Tensor \\($c_\alpha$, $c^\dagger_\alpha$)} & $\phi_{\alpha\beta}$ & $c^\da_\alpha c_\beta\phi_{\alpha\beta}$ & 
        \begin{minipage}{0.4\textwidth}
        \vspace*{10pt}
		\begin{flushleft} 
			$[c_\alpha]=1/4$, $\chi_{\rm imp}={\mc C}_{\rm cp}/T$ (cf. Fig.~\ref{TensorImpSusceptibility}). \\
			Impurity not fully screened.  
		\end{flushleft}
        \vspace*{3pt}
        \end{minipage}
        \\
		\hline
		
		% Tensor impurity and vector bulk. 
		\ref{TensorImpVecBulk} & \makecell{Traceless Sym. \\ $N_b$-Tensor \\ ($b_\alpha$, $b^\da_\alpha$)} & $\phi_{\alpha}$ & $b^\da_\alpha b_\beta \phi_\alpha\phi_\beta+\cdots$  &
        \begin{minipage}{0.4\textwidth}
        \vspace*{10pt}
		\begin{flushleft} 
		  $N=3$: for $S=1$ ($N_b=1$), a fully screened fixed point is conjectured to be present in the phase diagram, but for $S\ge 2$ this fixed point is not found. \\
		  $N\geq 4$: a fully screened fixed point is present in the phase diagram. \end{flushleft}
          \vspace*{3pt}
          \end{minipage} 
          \\
		\hline
		
	\end{tabular}
	\caption{Summary of the impurity models. The first column indicates the section numbers. The second column contains the $SO(N)$ symmetry representation of the impurity Hilbert space as well as the parton degrees of freedom. $\gamma$, $c$ and $b$ represent Majorana fermion, canonical complex fermion and canonical complex boson, respectively. Bulk fields are listed in the third column, either an $O(N)$ vector $\phi_\alpha$ or an antisymmetric tensor $\phi_{\alpha\beta}$. For the latter case, the bulk critical point being considered is fine tuned to have $O(N(N-1)/2)$ symmetry as explained in the main text. The fourth column shows the bulk-impurity coupling at $\bs r=0$, where repeated indices are summed over from $1$ to $N$. The low-energy property of these impurity problems are given in the last column, where $[\cdot]$ denotes the scaling dimension and $\chi_{\rm imp}$ is the impurity static susceptibility defined in \eqref{chiimpDef}. The scaling dimension results for the first three models are valid in the $N\rightarrow \infty$ limit. In the last model and when $N=3$, we actually allow the impurity to have an arbitrary spin quantum number $S\geq 1$, as explained in the main text. }
	\label{Table:Summary}
\end{table*}

We consider a spin-$1/2$ impurity coupled to a 2+1D gapless bulk described by the critical $O(3)$ non-linear $\sigma$-model (NL$\sigma$M). More precisely, we write down the following Euclidean action: 
\begin{align}
S&=S_{\rm b}+S_{\rm imp}+S_{\rm cp}, \label{O3TotalAction}\\
S_{\rm b}&=\frac{1}{2}\int \rmd^3 x\left[ \sum_{\alpha=1}^3(\pa_\mu\phi_\alpha)^2+\lambda(x) \left( \sum_{\alpha=1}^3 \phi_\alpha^2-\frac{3}{g}\right)\right], \label{O3Sb}\\
S_{\rm imp}&=\frac{1}{4}\int\rmd\tau\sum_{i=0}^3\gamma_i(\tau)\pa_\tau\gamma_i(\tau), \\
S_{\rm cp}&=J\int \rmd\tau \sum_{\alpha=1}^3\rmi\gamma_0(\tau)\gamma_\alpha(\tau)\phi_\alpha(\tau,{\bs r}=0). \label{O3Scp}
\end{align}
Here $\lambda(x)$ is a Lagrange multiplier field enforcing the NL$\sigma$M constraint $\phi_\alpha \phi_\alpha = \frac{3}{g}$, and $g$ is the bulk coupling constant. 
We use a Majorana representation of the spin impurity; there are four Majorana operators $\gamma_i~(i=0,1,2,3)$ acting on a four-dimensional extended Hilbert space and we project onto one of the two fermion-number parity sectors, say $P_f=-\gamma_0\gamma_1\gamma_2\gamma_3=1$ for concreteness. The spin operators which generate the $SO(3)$ symmetry on the impurity spin Hilbert space are represented as $S_{\alpha\beta}=-\frac{1}{2}\rmi\gamma_\alpha\gamma_\beta$ for $\alpha,\beta=1,2,3$ and, specially for the $SO(3)$ case, we define $S_\alpha=\frac{1}{2}\epsilon^{\alpha\beta\gamma}S_{\beta\gamma}\propto \gamma_0\gamma_\alpha$, e.g. $S_1=S_{23}\propto\gamma_0\gamma_1$. It is clear that $\gamma_\alpha$ and therefore $\gamma_0\gamma_\alpha$ for $\alpha=1,2,3$ transform under the vector representation of $SO(3)$. However, $\gamma_0\gamma_\alpha$ have to transform trivially under the $\mathbb{Z}_2$ inversion subgroup of the $O(3)$ symmetry since $\prod_{\alpha=1}^3(\gamma_0\gamma_\alpha)\propto P_f$ and we restrict to a certain fermion-number parity. As a result, the coupling term $S_{\rm cp}$ breaks the $O(3)$ symmetry down to $SO(3)$. It is important to note that the $SO(3)$ symmetry is represented projectively on the impurity. The two different signs of $J$ are equivalent by the field redefinition $\phi\mapsto-\phi$ which is a symmetry of the bulk action $S_{\rm b}$. 

This $SO(3)$ problem was first analyzed in $D=4-\epsilon$ spacetime dimensions in \cite{SachdevScience,SubirImp} for a general spin-$S$ impurity. They identified an interacting fixed point near the decoupled one for small $\epsilon$, where the impurity is not fully screened and exhibits a Curie form static susceptibility at low temperature. In this work, we study several $SO(N)$ generalizations of this problem in the large-$N$ limit. For comparison, we also study a slightly different model that enjoys the $\phi \mapsto -\phi$ symmetry of the bulk. 

Let $N\in 2\mathbb{Z}+1$. The most straightforward $SO(N)$ symmetric generalization of this impurity problem is to consider the critical $O(N)$ NL$\sigma$M in the bulk and an $SO(N)$ spinor impurity. More precisely, we have vector fields $\phi_\alpha~(\alpha=1,2,\cdots,N)$ in the bulk. On the impurity site, we introduce Majorana operators $\gamma_0$,  $\gamma_\alpha~(\alpha=1,2,\cdots,N)$ and project onto a certain fermion-number parity sector, say $P_f=(-\rmi)^{(N+1)/2}\gamma_0\gamma_1\cdots\gamma_N=1$. The most relevant coupling between the impurity and the bulk is proportional to $\rmi\gamma_0\gamma_\alpha\phi_\alpha(\bs r=0)$. It is obvious that this new model reduces to the original one at $N=3$. 

The above generalization, though simple, has one disadvantage: it is hard to include other types of impurity. For example, when $N=3$, we may instead consider a spin-$1$ impurity, but it does not have a direct generalization with bulk being the $O(N)$ NL$\sigma$M. Indeed, in order to couple to the vector fields $\phi_\alpha$, one needs impurity \emph{operators} which transform as an $SO(N)$ vector, but there is no such operator when the impurity \emph{Hilbert space} is in the vector representation. There is an alternative large-$N$ model where we introduce antisymmetric matrix fields in the bulk which transform in the $SO(N)$ adjoint representation, namely $\phi_{\alpha\beta}~(\alpha,\beta=1,2,\cdots,N)$ and $\phi^T=-\phi$. There are now $N(N-1)/2$ independent bulk $\phi$ fields and we require fine-tuning to the critical point with an emergent $O(N(N-1)/2)$ symmetry. On the impurity site, we may have the same $SO(N)$ spinor representation and the Majorana operators $\gamma_0,\gamma_\alpha$ subjected to a fermion-number parity projection. The most relevant bulk-impurity coupling is $\rmi\gamma_\alpha\gamma_\beta\phi_{\alpha\beta}(\bs r=0)$. It is straightforward to check that this new large-$N$ problem also reduces to the original one at $N=3$ since vector and adjoint representations are equivalent for $SO(3)$. This new bulk theory has the disadvantage of extra fine-tuning; one at least needs to turn off the single-trace term $\Tr(\phi^4)$ in the bulk. However, it enables us to consider impurities in different $SO(N)$ representations as motivated earlier. Indeed, $\phi_{\alpha\beta}$ can directly couple to the $so(N)$ generators which exist in any representation. 

With these motivations, in this paper, we consider the following three large-$N$ generalizations of the original $SO(3)$ problem:
\begin{itemize}
	\item[1.] An $SO(N)$ spinor impurity coupled to vector bulk fields $\phi_\alpha$. 
	
	\item[2A.] An $SO(N)$ spinor impurity coupled to adjoint bulk fields $\phi_{\alpha\beta}$. 
	
	\item[2B.] A totally antisymmetric $\nu N$-tensor impurity coupled to adjoint bulk fields. 
	
	%\item[3.] A totally symmetric $\nu N$-tensor impurity coupled to vector bulk fields. 
\end{itemize}
For all these three models, we find nontrivial fixed points where the impurity is not fully screened by the bulk and these can be explained by symmetry fractionalization arguments. To make a contrast with these intermediate-coupling fixed points, we also introduce the fourth model: 
\begin{itemize}
    \item[3.] An impurity in the $O(N)$ model respecting the $\phi\mapsto -\phi$ symmetry. 
\end{itemize}
Interestingly, this model exhibits very different behaviors for $N=3$ and $N\geq 4$. For $N=3$, we allow the impurity to be in an arbitrary spin-$S$ representation with $S\geq 1$ and find that the low-energy physics has a sensitive dependence on $S$. More precisely, when $S=1$, there is an IR stable fully screened fixed point, but when $S\geq 2$ (including integer $S$), the fully screened fixed point becomes inaccessible in our analytic approach, instead, we find a runaway flow to some yet unknown strong coupling fixed point. For $N\geq 4$, the impurity is taken to transform in the traceless symmetric tensor representation of $SO(N)$ with $N_b$ indices. Here we find a fully screened fixed point for all values of $N_b$. 

The implications of our results for $N=3$ are the following. When a spin-$1/2$ impurity is coupled to the $O(3)$ critical bulk, our findings for Models 1 and 2A imply the impurity to be not fully screened. If we instead consider a spin-$1$ impurity, we conclude from Models 2B and 3 that there exist both a stable not fully screened fixed point and a fully screened fixed point. 

In Section~\ref{DifferentApproaches}, we compare our findings with $\epsilon$-expansion and QMC results on impurity problems at $N=3$. In Section~\ref{SYKConnections}, we show that Model 2A and 2B are closely connected with the real and complex SYK models, respectively. We give a brief summary of our main results in Table~\ref{Table:Summary}.

\section{Spinor Impurity with Vector Bulk Fields}\label{SpinorImpVecBulk}
In this section, we consider an $SO(N)$ spinor impurity coupled to the critical $O(N)$ NL$\sigma$M, namely the following Euclidean action. 
\begin{align}
S&=S_{\rm b}+S_{\rm imp}+S_{\rm cp}, \label{Model1TotalAction}\\
S_{\rm b}&=\frac{1}{2}\int \rmd^3 x\left[ \sum_{\alpha=1}^N(\pa_\mu\phi_\alpha)^2+\lambda(x) \left( \sum_{\alpha=1}^N \phi_\alpha^2-\frac{N}{g}\right)\right], \label{VectorBulkAction}\\
S_{\rm imp}&=\frac{1}{4}\int\rmd\tau\sum_{i =0}^N\gamma_i(\tau)\pa_\tau\gamma_i(\tau), \\
S_{\rm cp}&=J\int \rmd\tau \sum_{\alpha=1}^N\rmi\gamma_0(\tau)\gamma_\alpha(\tau)\phi_\alpha(\tau,{\bs r}=0), \label{Model1Scp}
\end{align}
where $N$ is assumed to be odd. We use a Majorana representation of the impurity where a fermion-number parity projection is implied. This projection may seem troubling, but fortunately, it is in fact unnecessary for our purpose. There is a $\mathbb{Z}_2$ unitary transformation, $\gamma_0$ together with the $\phi\mapsto -\phi$ bulk symmetry operation, which flips the fermion-number parity but commutes with the Hamiltonian. As a consequence, the correlation function of any set of operators which have even total charge under this $\mathbb{Z}_2$ transformation takes exactly the same value in the two fermion-number parity sectors and there is no need to apply the projection; examples include the two-point functions of the spin operator $S_{\alpha\beta}\propto \gamma_\alpha\gamma_\beta$ or the $SO(N)$ vector operator $\rmi\gamma_0\gamma_\alpha$. We will therefore consider the extended problem where the fermion operators are physical since for the quantities of interest, the projection to the fixed fermion-number parity sector makes no difference. Notice that the sign of $J$ in $S_{\rm cp}$ is irrelevant since it can be flipped by the $\phi\mapsto -\phi$ bulk symmetry. The large-$N$ limit of the bulk action $S_{\rm b}$ is well-known \cite{ZinnJustinReview}. In particular, at zero temperature and when $g$ is slightly above the critical value $g_c$, the two-point correlation function of $\phi$ is given by 
\begin{align}
g_\phi(x)&\equiv\ex{T\phi_{\alpha}(x)\phi_{\alpha}(0)}~~~(\text{$\alpha$  arbitrary})\nonumber\\
&=\int\frac{\rmd\omega\rmd^2p}{(2\pi)^3}\frac{1}{\omega^2+p^2+m^2}\rme^{-\rmi\omega\tau+\rmi\bs p\cdot\bs r}, 
\end{align}
where $m=\xi^{-1}\sim g-g_c$ is the inverse correlation length. For later convenience, let us introduce some related notation. We define $g^\lambda_\phi(x,x')$ as a generalized two-point function whose matrix inverse is 
\begin{align}
(g^\lambda_\phi)^{-1}(x,x')=-\pa^2\delta(x-x')+\lambda(x)\delta(x-x'). 
\end{align}
In the special case $x=(\tau,0)$ and $x'=(\tau',0)$, we denote $g_\phi(x-x')$ and $g^\lambda_\phi(x,x')$ by $g_{\phi,0}(\tau-\tau')$ and $g^\lambda_{\phi,0}(\tau,\tau')$, respectively. 

Now we apply standard path integral manipulations to the action $S$, similar to the ones described in \cite{SUNKondo}. First we integrate out all $\phi$ fields in the bulk, leading to a nonlocal four-fermion interaction on the impurity. Next, we decouple this interaction into $\gamma_0\gamma_0$ and $\gamma_\alpha\gamma_\alpha$ channels using a Hubbard-Stratonovich transformation. This process introduces bosonic bilocal fields $Q(\tau,\tau')$ and $\bar Q(\tau,\tau')$ that are antisymmetric by definition. Finally, we integrate out all fermion fields. The impurity part of the final action reads
\begin{align}
	S'&=\frac{1}{2J^2}\int\rmd\tau\rmd\tau' \bar Q(\tau,\tau')\left[ g_{\phi,0}^\lambda(\tau,\tau') \right]^{-1}Q(\tau,\tau')\nonumber\\
	&-\ln\Pf\left[ \frac{1}{2}\pa_\tau \delta(\tau-\tau')-Q(\tau,\tau')\right]\nonumber\\
	&-N\ln\Pf\left[ \frac{1}{2}\pa_\tau \delta(\tau-\tau')-\bar Q(\tau,\tau')\right]. 
\end{align}
Here, $\left[g^\lambda_{\phi,0}(\tau,\tau')\right]^{-1}$ is the ordinary number inverse instead of the matrix inverse. There is also a bulk part of the action that we did not show and it has an overall $N$ factor. If we choose $J=J_0/\sqrt{N}$, then the first term contains a factor of $N$ which suppresses the fluctuations of $Q$ and $\bar Q$ in the large-$N$ limit and with a proper choice of the path integral contours. We thus expect the existence of a controlled large-$N$ limit where $Q$, $\bar Q$ and $\lambda$ all become classical and their expectation values can be solved from the saddle-point equations. 
We assume the time translation symmetry. Let $G_0(\tau)=\ex{T\gamma_0(\tau)\gamma_0(0)}$ and $G(\tau)=\ex{T\gamma_\alpha(\tau)\gamma_\alpha(0)}$ for an arbitrary $\alpha=1,\cdots,N$ be the Majorana Green's functions. As $N\rightarrow \infty$, we have
\begin{align}
	G^{-1}_0(\tau,\tau')&=\frac{1}{2}\pa_\tau \delta(\tau-\tau')-Q(\tau,\tau'),\\
	G^{-1}(\tau,\tau')&=\frac{1}{2}\pa_\tau \delta(\tau-\tau')-\bar Q(\tau,\tau'), 
\end{align}
where $G^{-1}_0(\tau,\tau')$ is the matrix inverse of $G_0(\tau,\tau')\equiv G_0(\tau-\tau')$, and similarly for $G^{-1}$. Anticipating that $G_0$ and $G$ are of order $1$, we have the following saddle-point equations for $Q$ and $\bar Q$ at infinite $N$: 
\begin{align}
	\bar Q(\tau,\tau')&=0,\\
	Q(\tau,\tau')&=J_0^2 g^\lambda_{\phi,0}(\tau,\tau')G(\tau-\tau'), 
\end{align}
where we used
\begin{align}
\frac{\delta \ln\Pf G^{-1}}{\delta \bar Q(\tau,\tau')}=\frac{1}{2}\Tr\left( G \frac{\delta G^{-1}}{\delta \bar Q(\tau,\tau')} \right)=G(\tau-\tau'), 
\end{align}
keeping in mind that $\bar Q(\tau,\tau')=-\bar Q(\tau',\tau)$ are not independent from each other. These equations imply that $Q$ is of order $1$ while $\bar Q$ is of order $1/N$. As a consequence, we see from the action $S'$ that the impurity has no order-$1$ correction to the bulk $\lambda(x)$; we can simply set $\lambda$ to its uniform bulk value $m^2$ and replace $g^\lambda_{\phi,0}$ by $g_{\phi,0}$. 

\subsection{Fermion correlation functions}
The fermion Green's functions can be easily solved from the self-consistent equations. We focus on the bulk critical point $g=g_c$. First consider the simplest situation $T=0$ where we set $m=0$. $\bar Q=0$ implies that $G(\tau)=\sign(\tau)$ is free. Using $g_{\phi,0}(\tau)=1/(4\pi|\tau|)$, we obtain the solution for $Q(\tau-\tau')\equiv Q(\tau,\tau')$: 
\begin{align}
	Q(\tau)=\frac{J_0^2}{4\pi\tau}\quad{\rm or}\quad Q(\rmi\omega)=\frac{1}{4}\rmi J_0^2\sign(\omega).   
\end{align}
The solution for $G_0$ is given by 
\begin{align}
	G_0(\rmi\omega)^{-1}=-\frac{1}{2}\rmi\omega-Q(\rmi\omega).
\end{align}
We are mostly interested in the long-time or small-frequency regime, namely $\tau^{-1}\ll \Lambda_{\rm UV}$ where $\Lambda_{\rm UV}$ is the UV energy scale including both $J_0^2$ and the bulk high-energy cutoff. Thus we may omit the $\rmi\omega/2$ term on the right-hand side of the above equation and get 
\begin{align}
	G_0(\tau)=\frac{4}{\pi J_0^2\tau}\quad (T=0,~\tau^{-1}\ll \Lambda_{\rm UV}). 
\end{align}
We have found that at the new fixed point, $\gamma_\alpha~(\alpha=1,\cdots,N)$ remains free (zero scaling dimension), while $\gamma_0$ acquires a scaling dimension $1/2$. One can easily compute the leading-order correlation functions of more fermions using Wick's theorem. In particular, the $SO(N)$ vector operators $\rmi\gamma_0\gamma_\alpha$ which directly couple to the $\phi$-fields in the bulk have a scaling dimension $1/2$, same as the leading-order dimension of $\phi_\alpha$. 

Next, consider nonzero temperature with $T,\tau^{-1}\ll\Lambda_{\rm UV}$. When $T>0$, a nonzero mass $m=\mu T$ is generated for the bulk $\phi$ bosons with $\mu=2\ln\left[(\sqrt{5}+1)/2\right]\approx 0.96$. Using 
\begin{align}
	g_{\phi,0}(\tau)&=\frac{1}{\beta}\sum_{\omega_n} \int\frac{\rmd^2p}{(2\pi)^2}\frac{1}{\omega^2_n+p^2+m^2}\rme^{-\rmi\omega_n\tau}\\
	&=\int\frac{\rmd^2p}{(2\pi)^2}\frac{\cosh\left[ \left(\frac{1}{2}\beta-|\tau|\right)\sqrt{p^2+m^2} \right]}{2\sqrt{p^2+m^2}\sinh\left( \frac{1}{2}\beta\sqrt{p^2+m^2} \right)},\label{gphi0} 
\end{align}
we found in the small frequency regime, 
\begin{align}
	G_0(\rmi\omega_n)=\frac{1}{J_0^2}\Phi_0(\rmi\tilde\omega_n), 
\end{align}
where 
\begin{align}
	\tilde\omega_n\equiv \omega_n/T,\quad \Phi_0(\rmi\tilde\omega_n)=-\frac{1}{\Psi(\rmi\tilde\omega_n)}, 
\end{align}
and 
\begin{align}
	\Psi(\rmi\tilde\omega_n)=\int_0^\infty\frac{\rmd x}{4\pi}~\frac{\rmi\tilde\omega_n\coth\left( \frac{1}{2}\sqrt{x+\mu^2} \right)}{\sqrt{x+\mu^2}\left( \tilde\omega_n^2+x+\mu^2 \right)}. 
\end{align}
Note that the dimensionless function $\Phi_0$ does not dependent on $J_0$ or the temperature $T$; one can also write in the time domain $G_0(\tau)=(T/J_0^2)\Phi_0(\tau T)$ with $\Phi_0(\rmi\tilde\omega_n)\equiv \int_0^1\rmd(\tau T)\rme^{\rmi\omega_n\tau}\Phi_0(\tau T)$. As $n\rightarrow\infty$, $\Phi_0(\rmi\tilde\omega_n)$ approaches its asymptotic form $4\rmi\sign(\tilde\omega_n)$ expected from the zero-temperature Green's function. $\gamma_\alpha$ with $\alpha\geq 1$ are still free with $G(\tau)=\sign(\tau)$ or $G(\rmi\omega_n)=2\rmi/\omega_n$. 

The equality between the scaling dimension of the $SO(N)$ vectors $\rmi\gamma_0\gamma_\alpha$ and that of the $\phi$ fields in the bulk seems to imply that the impurity completely merges into the bulk. However, this is not the case. One simple evidence is that the $so(N)$ generators $\rmi\gamma_\alpha\gamma_\beta~(\alpha,\beta\geq1)$ on the impurity site have zero scaling dimension. In the following, we will compute a physical response of the impurity showing that it is not fully screened by the bulk. 

\subsection{Impurity susceptibility}
The static susceptibility $\chi$ of the system to a uniform magnetic field in general takes the form: 
\begin{align}
\chi=\mc A\chi_{\rm b}+\chi_{\rm imp}, 
\label{chiimpDef}
\end{align}
where $\mc A$ is the spatial area. As the temperature $T$ goes to zero, the bulk susceptibility $\chi_{\rm b}$ goes to zero linearly in $T$ \cite{Chubukov&Sachdev}. However, as we will show below, the impurity susceptibility $\chi_{\rm imp}$ actually diverges as $T^{-1}$, which means the impurity spin does not merge into the bulk. 

Let us first determine how to couple a slowly-varying magnetic field to the system. In the $SO(3)$ case, the magnetic field has three components $H_\alpha(x)$ with $\alpha=1,2,3$ and microscopically couples to each spin degree of freedom $\mc S_\alpha$, either in the bulk or at the impurity site, as $-\sum_\alpha H_\alpha(x)\mc S_\alpha$. For the impurity site, we just replace $\mc S_\alpha$ by $S_\alpha$, and for the bulk spins, the correct modification of the action turns out to be the replacement \cite{SubirBook}
\begin{align}
\pa_\tau\phi_\alpha\mapsto\pa_\tau\phi_\alpha-\rmi(\bs H\times \bs \phi)_\alpha. 
\label{CouplingHtoBulkSO3}
\end{align}
In the $SO(N)$ case, we generalize the magnetic field to an antisymmetric matrix $H_{\alpha\beta}$ with $\alpha,\beta=1,\cdots,N$ which has $N(N-1)/2$ independent components and reduces to a vector field in the $SO(3)$ case by $H_\alpha=\frac{1}{2}\epsilon^{\alpha\beta\gamma}H_{\beta\gamma}$. The coupling of $H_{\alpha\beta}$ to the impurity site is 
\begin{align}
\Delta\mc L_{\rm imp}=\frac{1}{2}\sum_{\alpha<\beta}H_{\alpha\beta}(\rmi\gamma_\alpha\gamma_\beta),  
\end{align}
and the coupling to the bulk is 
\begin{align}
\pa_\tau\phi_{\alpha}\mapsto \pa_\tau\phi_{\alpha}+\rmi H_{\alpha\beta}\phi_\beta, 
\end{align}
both of which reduce to the correct form at $N=3$. 

To compute the susceptibility $\chi$, we take $H_{12}(x)=-H_{21}(x)=h$ to be a constant and turn off other components of $H_{\alpha\beta}$. Then $\chi$ is given by 
\begin{align}
\chi=T\left.\frac{\pa^2\ln Z}{\pa h^2}\right|_{h=0}=T\frac{1}{Z}\left.\frac{\pa^2Z}{\pa h^2}\right|_{h=0}, 
\end{align}
where we used $(\pa Z/\pa h)|_{h=0}=0$ since the $SO(N)$ symmetry can reverse the sign of $h$. The additional terms in the Lagrangian now take the form
\begin{align}
&\Delta\mc L_{\rm b}=2\rmi h(\pa_\tau\phi_{1})\phi_{2}-\frac{1}{2}h^2\left( \phi^2_{1}+\phi^2_{2} \right), \nonumber\\
&\Delta\mc L_{\rm imp}=\frac{1}{2}h(\rmi\gamma_1\gamma_2). 
\label{TermswH12_VecBulkSpinorImp}
\end{align}
First consider the decoupled system where $J_0=0$. It is straightforward to show that $\chi_{\rm imp}=1/(4T)\equiv \mc C_{\rm free}/T$ for all $N$. This coincides with the well-known result $S(S+1)/(3T)$ for a free $SO(3)$ spin with $S=1/2$. Next consider the much more complicated coupled system. Depending on whether $\pa/\pa h$ acts on $\Delta\mc L_{\rm b}$ or $\Delta\mc L_{\rm imp}$, the impurity susceptibility naturally separate into three terms: 
\begin{align}
\chi_{\rm imp}=\chi_{\rm b,b}+2\chi_{\rm b,imp}+\chi_{\rm imp,imp}. 
\label{chiimpDecomp}
\end{align}
These three terms can be computed order-by-order in $1/N$. At order-$1$, we found that only $\chi_{\rm imp,imp}$ contributes: 
\begin{align}
	\chi_{\rm imp,imp}^{(0)}=\frac{1}{4\beta}\sum_\omega G(\rmi\omega_n)G(-\rmi\omega_n)=\frac{1}{4T}. 
\end{align}
This is the same as a free spin, a not surprising result since we have found that $\gamma_\alpha$ with $\alpha\geq 1$ are free. We have thus shown that the impurity is not screened by the bulk; it is in fact free to the leading order. We can go to the next order to reveal a nontrivial interaction effect. 
\begin{figure}[h]
	\centering
	\includegraphics[width=1\linewidth]{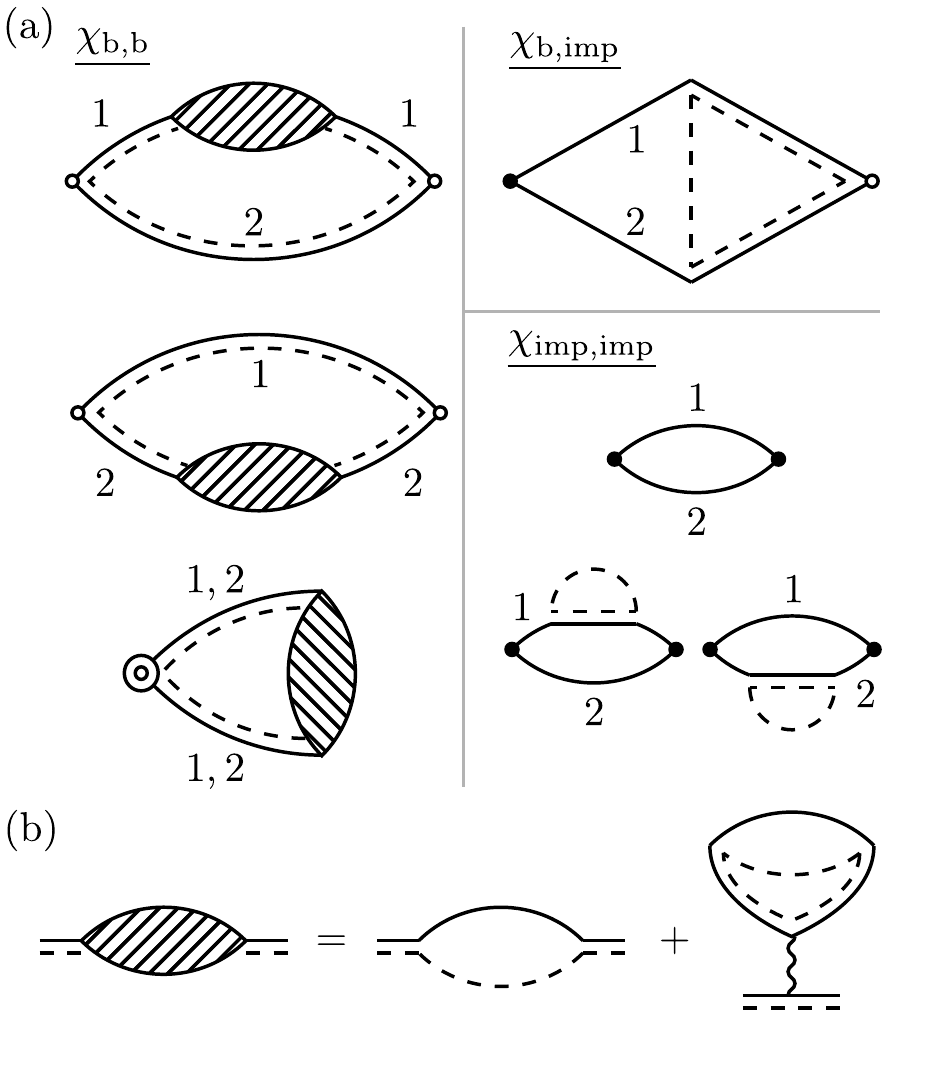}
	\caption{(a) Feynman diagrams at order $1$ and $1/N$ for the impurity susceptibility $\chi_{\rm imp}$. Single circles are bulk source terms of the form $(\pa_\tau\phi)\phi$, double circles are bulk source terms of the form $\phi^2$ which couples to $h^2$ and black solid dots are impurity source terms of the form $\gamma\gamma$ (cf. \eqnref{TermswH12_VecBulkSpinorImp}). 
	The solid lines are propagators for $\gamma_\alpha$ with $\alpha\geq 1$ and the dashed lines are $\gamma_0$ propagators. $\phi$ boson propagators are represented by double lines (solid and dashed).  
	(b) The shaded bubble which appears in the first panel. The wavy line represents the propagator of the Lagrange multiplier field $\lambda$ and is of order $1/N$. }
	\label{SusceptibilityDiagrams_VecBulk}
\end{figure}
\figref{SusceptibilityDiagrams_VecBulk}a presents all the diagrams of order $1$ and $1/N$ for $\chi_{\rm imp}$, where the shaded bubble is explained in panel b. Computing all the diagrams in \figref{SusceptibilityDiagrams_VecBulk}, we found that 
\begin{align}
\chi_{\rm imp}=\frac{\mc C_{\rm cp}}{T}. 
\end{align}
There is no anomalous dimension in $T$, but the Curie coefficient is renormalized, coinciding with the qualitative behavior obtained from $\epsilon$-expansion in \cite{SubirImp}. Let us avoid going into the full details of this calculation, but only note that there are remarkable divergence cancellations. One cancellation is between $\chi_{\rm imp,imp}$ and $2\chi_{\rm b,imp}$, both of which are logarithmically divergent. The other happens in the computation of the shaded bubble as depicted in \figref{SusceptibilityDiagrams_VecBulk}b and appearing in $\chi_{\rm b,b}$. There are two terms contributing to this shaded bubble, one is the fermion bubble and the other is the $1/N$ correction to the $\lambda$ expectation value. Both of them contain logarithmic UV divergences, but their sum is finite. The interaction effect of the bulk theory, more precisely the $\lambda\phi^2$ vertex, plays an important role for this latter cancellation. In \cite{SubirImp}, a similar fermion bubble divergence showed up in the $SU(N)$ large-$N$ approach but since the bulk there was assumed to be free, this divergence remained un-cancelled and would result in a log-divergence of the susceptibility. In fact, the model in Ref. \cite{SubirImp} is  more closely related to the case of tensor bulk fields to be discussed later. The analytic expressions for $\mc C_{\rm cp}$ can be found in Appendix~\ref{SusceptibilityExpression}. Numerically we obtained 
\begin{align}
\mc C_{\rm cp}=\left[ 1+\frac{1}{N}\left( -0.4187\pm 9\times 10^{-5} \right) \right]\mc C_{\rm free}, 
\label{chiResult_SpinorImpVecBulk}
\end{align}
where $\mc C_{\rm free}=1/4$ has been defined above. If one plugs in $N=3$, this gives $\mc C_{\rm cp}=0.8604~\mc C_{\rm free}$. Numerically, infinite frequency sums are truncated by some maximal frequency $\tilde\omega_{\rm max}$. The error above comes from an extrapolation to $\tilde\omega_{\rm max}=\infty$. 

The impurity being not fully screened is a consequence of symmetry fractionalization. The $SO(N)$ symmetry in the bulk fractionalizes into a $Spin(N)$ symmetry on the impurity site, therefore the bulk does not have the right degrees of freedom to fully screen the impurity. More explicitly, the counterclockwise spin rotation by $\theta$ on the $\alpha\beta$-plane is represented on the impurity Hilbert space as $R_{\alpha\beta}(\theta)=\exp(-\rmi\theta S_{\alpha\beta})=\cos(\theta/2)-\sin(\theta/2)\gamma_\alpha\gamma_\beta$, where we used $S_{\alpha\beta}=-\frac{1}{2}\rmi\gamma_\alpha\gamma_\beta$. We see that the $2\pi$ rotation (on an arbitrary plane) acts as $-1$ on the impurity. On the contrary, a $2\pi$ rotation is equivalent to the identity in the $SO(N)$ group, for example it acts trivially on the bulk fields $\phi_\alpha$. Hence the impurity spins furnish a projective representation of $SO(N)$. 

\section{Spinor Impurity with Tensor Bulk Fields}\label{SpinorImpTensorBulk}
In this section, we consider an alternative generalization of the $N=3$ problem described earlier in Section~\ref{SO3Problem}. We demand the bulk scalar fields to transform under the \emph{adjoint} representation of $SO(N)$, instead of the vector representation. Note that these two are equivalent in the special case $N=3$. We therefore introduce an antisymmetric matrix field $\phi_{\alpha\beta}~(\alpha,\beta=1,\cdots,N)$ which satisfies $\phi_{\alpha\beta}=-\phi_{\beta\alpha}$ and thus has $N_{\rm b}\equiv N(N-1)/2$ number of independent components. The bulk action is given by 
\begin{align}
	S_{\rm b}&=\frac{1}{2}\int \rmd^3 x\left[ \sum_{\alpha<\beta}(\pa_\mu\phi_{\alpha\beta})^2+\lambda(x) \left( \sum_{\alpha<\beta} \phi_{\alpha\beta}^2-\frac{N_{\rm b}}{g}\right)\right], \label{TensorBulkAction}
\end{align}
which has an emergent $O(N(N-1)/2)$ symmetry containing the original $O(N)$ as a subgroup. The large-$N$ limit of this action has been explained in Section~\ref{SpinorImpVecBulk} and we will be using the same notations. As before, we take $N$ to be odd. 

For the spin impurity, we again introduce $N+1$ Majorana operators $\gamma_i~(i=0,\dots,N)$ acting on a $2^{(N+1)/2}$-dimensional extended Hilbert space and then project onto one of the two fermion-number parity sectors, say $P_f=(-\rmi)^{(N+1)/2}\gamma_0\gamma_1\cdots\gamma_N=1$. The impurity and the coupling terms in the action take the form
\begin{align}
	S_{\rm imp}+S_{\rm cp}&=\frac{1}{4}\int\rmd\tau\sum_{i=0}^N\gamma_i\pa_\tau\gamma_i\nonumber\\
	&+J\int\rmd\tau\sum_{\alpha<\beta}(\rmi\gamma_\alpha\gamma_\beta)\phi_{\alpha\beta}(\tau,{\bs r}=0), \label{Model2ASimpScp}
\end{align}
where we have ignored $\gamma_0$ which completely decouples from other fields. The sign of $J$ is irrelevant as before due to the $\phi\mapsto-\phi$ bulk symmetry. Similar to Section~\ref{SpinorImpVecBulk}, there is no need to actually worry about the fermion-number parity projection: the operator $\gamma_0$ serves as a unitary transformation that flips the fermion-number parity but commutes with the Hamiltonian, which guarantees all correlation functions that we will be interested in take the same value in the two fermion-number parity sectors. We will therefore again consider the extended problem where the fermion operators are physical. 

Now we analyze the large-$N$ limit of the coupled system. Integrating out all the $\phi$ fields in the bulk, then decoupling the resulting four-fermion interaction term into $\gamma_\alpha\gamma_\alpha$ channels by an antisymmetric bilocal field $Q(\tau,\tau')$, and finally integrating out all fermion fields, we obtain the following impurity part of the action:  
\begin{align}
	S'&=-N\ln\Pf\left[ \frac{1}{2}\pa_\tau \delta(\tau-\tau')-Q(\tau,\tau')\right]\nonumber\\
	&+\frac{1}{4J^2}\int\rmd\tau\rmd\tau'Q^2(\tau,\tau')(g^\lambda_{\phi,0}(\tau,\tau'))^{-1}. 
\end{align}
We now choose $J^2$ to scale as $J^2_0/N$ and find that $S'$ contains an overall factor of $N$. Moreover, the bulk effective action that we did not show explicitly has an $N_{\rm b}$ factor, thus as $N\rightarrow\infty$, both $\lambda(x)$ and $Q(\tau,\tau')$ become classical and their expectation values can be obtained from the saddle-point equations. Since $N_{\rm b}$ is of order $N^2$, the classical solution of $\lambda$ is the same as that in the absence of impurity up to order $1/N$ corrections. As a result, the classical solution for $Q(\tau,\tau')$ can be obtained by setting $\lambda(x)=m^2$. Let $G_\gamma(\tau)=\ex{T\gamma_\alpha(\tau)\gamma_\alpha(0)}$ with an arbitrary $\alpha=1,\cdots,N$ be the Majorana Green's function. In the large-$N$ limit where $Q(\tau,\tau')$ is pinned to its saddle-point solution, we have 
\begin{align}
	&G^{-1}_\gamma(\tau,\tau')=\frac{1}{2}\pa_\tau \delta(\tau-\tau')-Q(\tau,\tau'). 
	\label{SelfConsistentEq1_0D}
\end{align}
The saddle-point equation for $Q(\tau,\tau')$ as $N\rightarrow\infty$ is 
\begin{align}
	Q(\tau,\tau')=J^2_0G_\gamma(\tau-\tau')g_{\phi,0}(\tau-\tau'),  
	\label{SelfConsistentEq2_0D}
\end{align}
where we have used
\begin{align}
	\frac{\delta \ln\Pf G^{-1}_\gamma}{\delta Q(\tau,\tau')}=\frac{1}{2}\Tr\left( G_\gamma \frac{\delta G^{-1}_\gamma}{\delta Q(\tau,\tau')} \right)=G_\gamma(\tau-\tau'). 
\end{align}

\subsection{Impurity spin correlation function}
The impurity spin Green's function $G_S(\tau)=\ex{T S_{\alpha\beta}(\tau)S_{\alpha\beta}(0)}$ (no sum over $\alpha$, $\beta$) with the definition $S_{\alpha\beta}=-\frac{1}{2}\rmi\gamma_\alpha\gamma_\beta$ can be computed as $G_S(\tau)=\frac{1}{4}G^2_\gamma(\tau)$ for $N \to \infty$. Our task now is to solve the self-consistent equations \eqref{SelfConsistentEq1_0D} and \eqref{SelfConsistentEq2_0D}. 
We focus on the bulk critical point $g=g_c$. First consider the situation $T=0$ and $\tau^{-1}\ll \Lambda_{\rm UV}$. Dropping the first term on the right hand side of Eq.\,(\ref{SelfConsistentEq1_0D}) and using $g_{\phi,0}(\tau)=1/(4\pi|\tau|)$, the self-consistent equations are solved with the ansatz
\begin{align}
	&G_\gamma(\tau)=\frac{A_\gamma}{|\tau|^{2\Delta}}\sign(\tau)\nonumber\\
	\Rightarrow~&G_\gamma(\rmi\omega)=\int_{-\infty}^\infty\rmd\tau\rme^{\rmi\omega\tau}G_\gamma(\tau)\nonumber\\
	&=2\rmi A_\gamma\cos(\pi\Delta)\Gamma(1-2\Delta)|\omega|^{2\Delta-1}\sign(\omega).  
\end{align}
% where the Fourier transform result holds for $0\leq\Delta<1/2$. 
when $\Delta=1/4$ and $A_\gamma^2=1/J_0^2$. {This scaling dimension of the Majorana fermions is the same as that in the SYK model with $q=4$ \cite{SYKRemarks} a point that we will return to in Section  \ref{SYKConnections}} below. 

It is easy to see from the definition\footnote{Recall that we are already considering an extended problem where fermion operators are physical. } that for all temperature, $G_\gamma(\tau)\geq0$ when $\tau\geq0$, thus $A_\gamma=1/|J_0|$. Notice that since $Q(\omega) \sim \sqrt{\omega}$, the term we dropped  in Eq.\,(\ref{SelfConsistentEq1_0D}) is indeed negligible. The spin two-point function is determined as 
\begin{align}
	G_S(\tau)=\frac{1}{4J_0^2|\tau|}~~~(T=0,~\tau^{-1}\ll \Lambda_{\rm UV}).  
\end{align}

\begin{figure}%[h]
	\centering
	\includegraphics[width=0.7\linewidth]{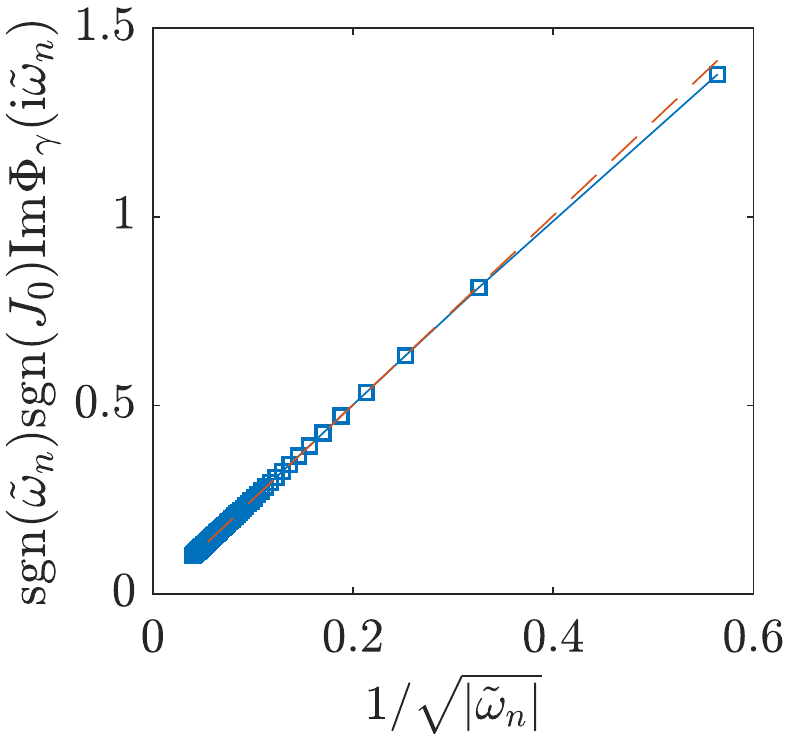}
	\caption{$\sign(\tilde\omega_n)\sign(J_0)\imag\Phi_\gamma(\mathrm{i}\tilde\omega_n)$ as a function of $1/\sqrt{|\tilde\omega_n|}$ for $100$ values of $|\tilde\omega_n|$. At large $|n|$, $\Phi_\gamma(\mathrm{i}\tilde\omega_n)$ approaches the zero-temperature form $\rmi\sign(J_0)\sign(\tilde\omega_n)\sqrt{2\pi}|\tilde\omega_n|^{-1/2}$, indicated by the red dashed line. }
	\label{ImpCorrelatorFiniteTem}
\end{figure}
Next, consider nonzero temperature with $T,\tau^{-1}\ll\Lambda_{\rm UV}$. Recall when $T>0$, a nonzero mass $m=\mu T$ is generated for the bulk $\phi$ bosons. We can write $G_\gamma(\tau)$ in the following scaling form: 
\begin{align}
	J_0G_\gamma(\tau)=\sqrt{T}\Phi_\gamma(\tau T),  
\end{align}
where $\Phi_\gamma$ is some universal function. In the frequency space, this becomes
\begin{align}
	J_0G_\gamma(\rmi\omega_n)=\frac{1}{\sqrt{T}}\Phi_\gamma(\rmi\tilde\omega_n), 
\end{align}
where we have used dimensionless frequencies $\tilde\omega_n=\omega_n/T=(2n-1)\pi$ and the corresponding Fourier transform $\Phi_\gamma(\rmi\tilde\omega_n)=\int_0^1\rmd(\tau T)\rme^{\rmi\tilde\omega_n(\tau T)}\Phi_\gamma(\tau T)$. 
$G_\gamma(\tau)$ being real and antisymmetric implies $G_\gamma(-\rmi\omega_n)=G_\gamma(\rmi\omega_n)^*$ as well as $G_\gamma(-\rmi\omega_n)=-G_\gamma(\rmi\omega_n)$, thus $\Phi_\gamma(\rmi\tilde\omega_n)$ is purely imaginary and odd in frequency. In terms of $\imag\Phi_\gamma(\rmi\tilde\omega_n)$, the self-consistent equations are equivalent to 
\begin{align}
	&\frac{1}{\imag\Phi_\gamma(\rmi\tilde\omega_n)}\nonumber\\&=-\frac{1}{4\pi}\sum_{\tilde\nu_k>0}\imag\Phi_\gamma(\rmi\tilde\nu_k)\ln\left[ \frac{(\tilde\omega_n-\tilde\nu_k)^2+\mu^2}{(\tilde\omega_n+\tilde\nu_k)^2+\mu^2} \right], 
\end{align}
where we have used 
\begin{align}
	g_{\phi,0}(\rmi\omega_n)=\frac{1}{4\pi}\ln\left( \frac{\Lambda^2}{\omega_n^2+m^2} \right)
	\label{gphi0Expression}
\end{align}
with $\Lambda$ - the UV cutoff of the bulk momentum. This equation can be solved numerically. In practice, to accurately estimate $\Phi_\gamma(\rmi\tilde\omega_n)$ up to some $\tilde\omega_{\rm max}=(2n_{\rm max}-1)\pi$, we need to truncate the frequency space at a larger $\tilde\omega_{\rm ext}=(2n_{\rm ext}-1)\pi$. As $|n|\rightarrow\infty$, we expect $\Phi_\gamma(\mathrm{i}\tilde\omega_n)$ to approach the zero-temperature form $\rmi\sign(J_0)\sign(\tilde\omega_n)\sqrt{2\pi}|\tilde\omega_n|^{-1/2}$. Using this asymptotic behavior, one can show that if we send both $\tilde\omega_{\rm max}$ and $\tilde\omega_{\rm ext}$ to infinity but have $\tilde\omega_{\rm max}/\tilde\omega_{\rm ext}\rightarrow 0$, then the relative error of $\Phi_\gamma(\mathrm{i}\tilde\omega_{\rm max})$ goes to zero. Therefore, we took $n_{\rm ext}=n_{\rm max}^2$ in our actual calculation. The result for $\Phi_\gamma(\rmi\tilde\omega_n)$ is shown in \figref{ImpCorrelatorFiniteTem}. It is clear that as $|n|$ gets large, $\Phi_\gamma(\mathrm{i}\tilde\omega_n)$ indeed approaches the expected zero-temperature form. 

\subsection{Impurity susceptibility}
Next, we compute the impurity susceptibility. We again generalize the magnetic field to an antisymmetric matrix $H_{\alpha\beta}$ with $\alpha,\beta=1,\cdots,N$ which reduces to a vector field in the $SO(3)$ case by $H_\alpha=\frac{1}{2}\epsilon^{\alpha\beta\gamma}H_{\beta\gamma}$. Recall $\phi_{\alpha\beta}$ also embeds into an antisymmetric matrix $\phi$ by  $\phi_{\alpha\beta}=-\phi_{\beta\alpha}$. The coupling of $H_{\alpha\beta}$ to the impurity site is 
\begin{align}
	\Delta\mc L_{\rm imp}=\frac{1}{2}\sum_{\alpha<\beta}H_{\alpha\beta}(\rmi\gamma_\alpha\gamma_\beta),  
\end{align}
and the coupling to the bulk is 
\begin{align}
	\pa_\tau\phi_{\alpha\beta}\mapsto \pa_\tau\phi_{\alpha\beta}-\rmi(\phi_{\alpha\gamma} H_{\gamma\beta}-H_{\alpha\gamma}\phi_{\gamma\beta}) 
\end{align}
which reduces to \eqnref{CouplingHtoBulkSO3} as one can easily check. As before, to compute the susceptibility $\chi$, we take $H_{12}(x)=-H_{21}(x)=h$ to be a constant and turn off other components of $H_{\alpha\beta}$. Then $\chi$ is given by 
\begin{align}
	\chi=T\left.\frac{\pa^2\ln Z}{\pa h^2}\right|_{h=0}=T\frac{1}{Z}\left.\frac{\pa^2Z}{\pa h^2}\right|_{h=0}.  
\end{align}
The additional terms in the Lagrangian now take the form
\begin{align}
	&\Delta\mc L_{\rm b}=h\sum_{\alpha>2}2(\rmi\pa_\tau\phi_{1\alpha})\phi_{2\alpha}-\frac{1}{2}h^2\sum_{\alpha>2}\left( \phi^2_{1\alpha}+\phi^2_{2\alpha} \right), \nonumber\\
	&\Delta\mc L_{\rm imp}=\frac{1}{2}h(\rmi\gamma_1\gamma_2). 
	\label{LAdditionalTermswH12}
\end{align}
In the decoupled limit $J_0=0$, we still have $\chi_{\rm imp}=1/(4T)\equiv \mc C_{\rm free}/T$ for all $N$. For the interacting system, it turns out that the low temperature behavior of $\chi_{\rm imp}$ is dominated by $\chi_{\rm b,b}$ (cf. \eqnref{chiimpDecomp}), thus we will focus on this term in the following. 
\begin{figure}[h]
	\centering
	\includegraphics[width=1\linewidth]{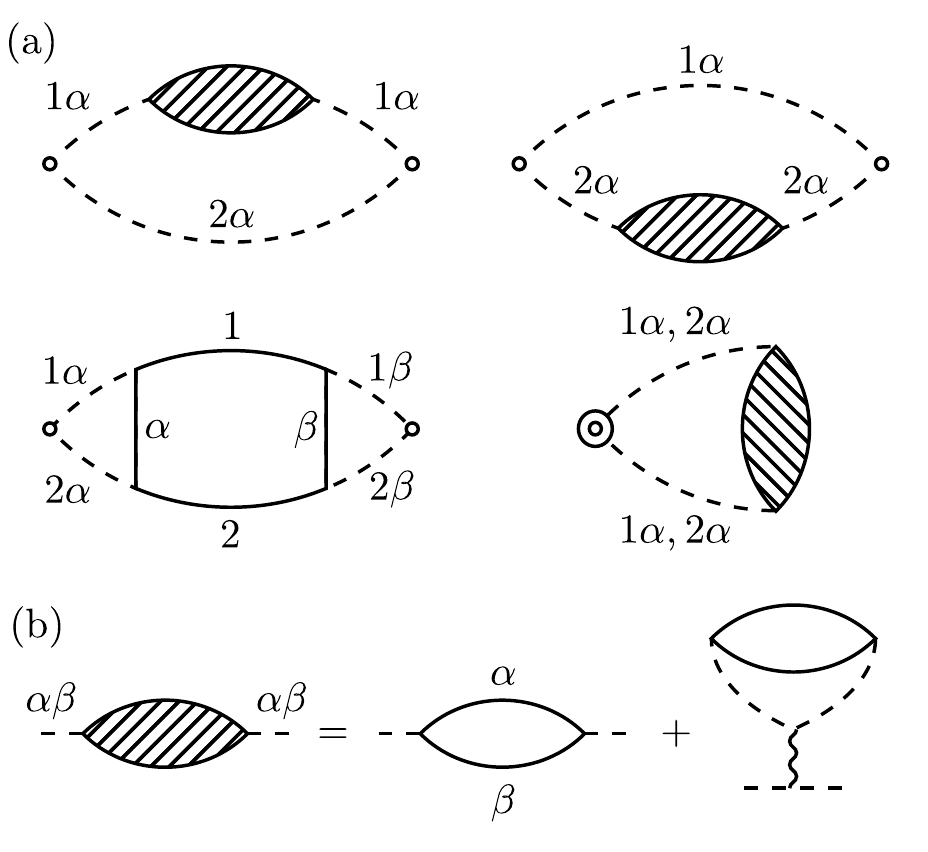}
	\caption{(a) Feynman diagrams relevant in the large-$N$ limit for $\chi_{\rm b,b}$, the bulk-bulk part of the impurity susceptibility to a uniform magnetic field. Single circles are source terms of the form $(\pa_\tau\phi)\phi$ which couples linearly to $h$, and double circles are sources terms of the form $\phi^2$ which couples to $h^2$ (cf. \eqnref{LAdditionalTermswH12}). The solid lines are Majorana fermion propagators and the dashed lines are $\phi$ boson propagators. (b) The shaded bubble which appears in the first panel. The wavy line represents the propagator of the Lagrange multiplier field $\lambda$ and is of order $1/N^2$. }
	\label{SusceptibilityDiagrams}
\end{figure}
\figref{SusceptibilityDiagrams}a presents all the diagrams of order $N^0$ for $\chi_{\rm b,b}$, where the shaded bubble is explained in panel b. We note that the two terms contributing to the shaded bubble again have a remarkable divergence cancellation. Computing all the diagrams in \figref{SusceptibilityDiagrams}, we found that 
\begin{align}
	\chi_{\rm imp}=\frac{\mc C_{\rm cp}}{T}. 
\end{align}
There is no anomalous dimension in $T$. The analytic expression for $\mc C_{\rm cp}$ is given in Appendix~\ref{SusceptibilityExpression}. Using the numerical solution for $\Phi_\gamma$, we found
\begin{align}
	\mc C_{\rm cp}=0.2109\pm 0.0018=(0.844\pm 0.007)\mc C_{\rm free}. 
	\label{chiResult_SpinorImpTensorBulk}
\end{align}
One may recall that in Section~\ref{SpinorImpVecBulk} (Model 1), the coupling to the bulk gives an order $1/N$ correction to the impurity Curie coefficient, in contrast to the order $1$ correction here. In fact, we now have an order $N^2$ number of fields in the bulk, and the bulk susceptibility scales as $N$ (in contrast to Model 1 where the bulk susceptibility is of $O(1)$). Thus, the  relative order in $N$ of boundary susceptibility correction to the bulk susceptibility is the same in Models 1 and 2. Notice that $\mc C_{\rm cp}$ in Eq.~(\ref{chiResult_SpinorImpTensorBulk}) is numerically quite close to the $N=3$ result in Section~\ref{SpinorImpVecBulk}, which is expected because the two models reduce to the same one at $N=3$. As in section \ref{SpinorImpVecBulk}, there is a similar symmetry fractionalization argument for the impurity being not fully screened. Let us briefly explain our numerical computation. Given a numerical solution of $\Phi_\gamma(\rmi\tilde\omega_n)$ up to $|\tilde\omega_n|\leq\tilde\omega_{\rm max}$, we can further extrapolate this solution by its known asymptotic form at larger frequencies. By increasing the number of frequency modes in the summation, we asymptotically converge to the susceptibility result corresponding to the infinite sum over frequency. 
%Given a numerical solution of $\Phi_\gamma(\rmi\tilde\omega_n)$ up to $|\tilde\omega_n|\leq\tilde\omega_{\rm max}$, we can further extend this solution by its known asymptotic form at large frequency. By varying the size of this extension, we extrapolate the obtained susceptibility results to the case of infinite frequency space. 
We collect two sources of error for the susceptibility obtained this way: one is from the extrapolation fitting, and the other is from the inaccuracy of $\Phi_\gamma(\rmi\tilde\omega_n)$ with $|\tilde\omega_n|\leq\tilde\omega_{\rm max}$. To estimate the latter error, we reduce $n_{\rm max}$ by half, compute the susceptibility in the same way and then evaluate the discrepancy. We regard the two kinds of errors as independent and the total error is the root mean square of them. For the numbers given above, we used $n_{\rm max}=100$, and the error is dominated by the extrapolation. 

\section{Antisymmetric Tensor Impurity with Tensor Bulk Fields}\label{TensorImpTensorBulk}
It is natural to generalize the previous analysis to impurities in other representations. For example, in the $SO(3)$ case, one may ask whether there is an even-odd effect; what will happen if we replace the spin-$1/2$ impurity (spinor representation) by spin-$1$ (vector representation)? For the $SO(N)$ case, we consider a totally antisymmetric tensor representation as explained below. 

We introduce complex fermion modes $c_\alpha$ for $\alpha=1,2,\cdots,N$, and make no assumption on whether $N$ is even or odd. The impurity spin operators $S_{\alpha\beta}$ are $-\rmi (c^\da_\alpha c_\beta-c^\da_\beta c_\alpha)$. We also fix the total fermion number as $\sum_\alpha c^\da_\alpha c_\alpha=\nu N$, namely the impurity Hilbert space is spanned by the states $c^\da_{\alpha_1}c^\da_{\alpha_2}\cdots c^\da_{\alpha_{\nu N}}\ket{0}$. When $\nu=1/N$, the impurity forms a vector representation of $SO(N)$. For a generic $\nu$, this is the antisymmetric $\nu N$-tensor representation. In the following, we will hold $\nu$ as a constant when taking the large-$N$ limit. We will still use a Majorana representation that manifest the analogy with the previous section: 
\begin{align}
	\gamma_{A\alpha}=c_\alpha+c^\da_\alpha,~\gamma_{B\alpha}=-\rmi(c_\alpha-c^\da_\alpha). 
\end{align}
The bulk consists of tensor fields $\phi_{\alpha\beta}$ with the action in \eqnref{TensorBulkAction}. The impurity and coupling terms in the action now take the form
\begin{align}
	S_{\rm imp}&=\frac{1}{4}\int\rmd\tau\sum_\alpha\sum_{I=A,B}\gamma_{I\alpha}\pa_\tau\gamma_{I\alpha}\\
	&+\int\rmd\tau~\xi(\tau)\left[ \sum_\alpha \rmi\gamma_{A\alpha}\gamma_{B\alpha}-(2\nu-1)N \right],\\
	S_{\rm cp}&=\sum_{I=A,B}\frac{J_0}{\sqrt{N}}\int\rmd\tau\sum_{\alpha<\beta}(\rmi\gamma_{I\alpha}\gamma_{I\beta})\phi_{\alpha\beta}(\tau,\bs r=0),
\end{align}
where $\xi(\tau)$ is a Lagrange multiplier field imposing the fermion-number constraint $\sum_\alpha \rmi\gamma_{A\alpha}\gamma_{B\alpha}=(2\nu-1)N$. We define the imaginary-time Green's functions\footnote{Strictly speaking, these correlation functions are ill-defined since the Majorana operators are fictitious. They should really be thought of as some bilocal fields in the path integral language and we are interested in their saddle point values from which we can compute physical correlation functions. } $G_{IJ}(\tau,\tau')=\ex{T\gamma_{I\alpha}(\tau)\gamma_{J\alpha}(\tau')}$ where $I,J\in\{ A,B\}$ and $\alpha$ is arbitrary. It directly follows from the definition that $G_{BA}(\tau,\tau')=-G_{AB}(\tau',\tau)$. For convenience, we will denote $G_{AA}$ and $G_{BB}$ by $G_A$ and $G_B$, respectively. We found the following self-consistent equations in the $N\rightarrow \infty$ limit: 
\begin{align}
	Q_I(\tau,\tau')&=J_0^2 g_{\phi,0}(\tau-\tau')G_I(\tau,\tau'),\label{TensorRepQIEq}\\
	R(\tau,\tau')&=J_0^2 g_{\phi,0}(\tau-\tau')G_{AB}(\tau,\tau'),\label{TensorRepREq}\\
	\rmi G_{AB}(\tau,\tau)&=2\nu-1, \label{FermionNumEq} 
\end{align}
In terms of the bilocal fields $Q_I$ and $R$, the Green's function matrix $G\equiv (G_{IJ})$ is given by
\begin{widetext}
\begin{align}
	G^{-1}(\tau,\tau')=
	\begin{pmatrix}
	\frac{1}{2}\pa_\tau\delta(\tau-\tau')-Q_A(\tau,\tau') & \rmi\xi(\tau)\delta(\tau-\tau')-R(\tau,\tau')\\
	-\rmi\xi(\tau)\delta(\tau-\tau')+R(\tau',\tau) & \frac{1}{2}\pa_\tau\delta(\tau-\tau')-Q_B(\tau,\tau')
	\end{pmatrix}
	\equiv M_\gamma(\tau,\tau'). 
	\label{Eq:Ginv}
\end{align}
\end{widetext}
In this expression, $G^{-1}$ is the inverse on both the $\{A,B\}$ and time indices. In other words, $\sum_J\int\rmd\tau' G_{IJ}(\tau,\tau')M_\gamma(\tau',\tau'')_{JK}=\delta_{IK}\delta(\tau-\tau'')$. $Q_A$ and $Q_B$ are by definition antisymmetric, but $R$ has no such restriction, i.e. $R(\tau,\tau')$ are all independent in the path integral. From these self-consistent equations, we look for solutions to the classical fields $Q_I$, $R$ and $\xi$. Our approach presented below is largely inspired by \cite{SUNKondo}. 
\subsection{Scaling ansatz}\label{TensorImpScalingAnsatzSubSec}
Before writing down the explicit ansatz for the Green's functions, let us first examine the general constraints they need to satisfy. The problem has a $U(1)$ \emph{gauge} symmetry generated by $\sum_\alpha\rmi\gamma_{A\alpha}\gamma_{B\alpha}$ which acts on Majorana operators as \begin{align}
	\begin{pmatrix}
	\gamma_{A\alpha}\\
	\gamma_{B\alpha}
	\end{pmatrix}
	\mapsto
	\begin{pmatrix}
	\cos\theta & -\sin\theta\\
	\sin\theta & \cos\theta
	\end{pmatrix}
	\begin{pmatrix}
	\gamma_{A\alpha}\\
	\gamma_{B\alpha}
	\end{pmatrix}. 
\end{align}
Equivalently, the action is $c_\alpha\mapsto e^{\rmi\theta}c_\alpha$. For now let us restrict to saddle point solutions which preserve this gauge symmetry and we will comment on other possibilities later. Taking $\theta=\pi/2$, we find $G_A(\tau,\tau')=G_B(\tau,\tau')$ and $G_{AB}(\tau,\tau')=-G_{BA}(\tau,\tau')\equiv G_{AB}(\tau',\tau)$. This implies $G_{AB}$ is symmetric. In contrast, $G_A$ is antisymmetric directly from the definition. We assume the time translation symmetry, thus $G(\tau,\tau')=G(\tau-\tau')$ and $\xi$ in Eq.\,(\ref{Eq:Ginv}) is time-independent. 

If we regard the $\gamma$'s as actual fermion operators, there will be more constraints on the Green's functions. On the one hand, $G_A(\tau)\in\mathbb{R}$ and $G_A(\tau)\geq0$ when $\tau>0$. On the other hand, $G_{AB}(\tau)^*=G_{BA}(\tau)=-G_{AB}(\tau)$, thus $G_{AB}$ is purely imaginary. It then follows from the expression for $G^{-1}$ that the saddle-point value of $\xi$ is real, thus it effectively sets a chemical potential. In the following, we will assume an ansatz with these properties. Of course there could be other solutions but there should be at least one solution satisfying these properties. This is because we can imagine really solving a problem of fermions on the impurity site with a chemical potential determined by $\xi$. As $N$ goes to infinity, we expect that the total fermion number does not fluctuate and therefore a solution for this problem also serves as a valid solution to our original problem a with fermion number projection. 

Define a self-energy matrix 
\begin{align}
Q(\tau)=
\begin{pmatrix}
Q_A(\tau) & R(\tau)\\
-R(\tau) & Q_A(\tau)
\end{pmatrix}, 
\end{align}
then the self-consistent equations can be compactly written as 
\begin{align}
[G(\rmi\omega_n)]^{-1}&=-\frac{1}{2}\rmi\omega_n-\xi \sigma_y-Q(\rmi\omega_n), \label{TensorImpCompactEqGInv}\\
Q(\tau)&=J_0^2g_{\phi,0}(\tau)G(\tau), \label{TensorImpCompactEqQ}
\end{align}
where $[G(\rmi\omega_n)]^{-1}$ denotes the $2\times 2$ matrix inverse and the symmetry constraints mentioned above have been used. The first equation is written in the frequency space with a nonzero-temperature form. When considering zero temperature, one just needs to replace $\omega_n$ by a continuous variable $\omega$. 

Now let us state our ansatz for the Green's functions at $T=0$. We emphasize that we only attempt to solve the self-consistent equations in the regime $\tau^{-1}\ll \Lambda$. In this long time limit, we expect the leading behavior of $G_A$ and $G_{AB}$ to be the following power-law form: 
\begin{align}
	G_A(\tau)&=\frac{A}{|\tau|^{2\Delta}}\sign(\tau)~~~(A\geq 0), \\
	G_{AB}(\tau)&=\frac{\rmi B}{|\tau|^{2\Delta}}~~~(B\in\mathbb{R}),  
\end{align}
where $A$ and $B$ are constants to be determined later. In terms of the complex fermion operators, \begin{align}
	\ex{T c_\alpha(\tau) c^\da_\alpha(0)}=\frac{A\sign(\tau)+B}{2|\tau|^{2\Delta}}\quad (\text{no sum over $\alpha$}). 
\end{align}
We assume $g_{\phi,0}$ at long time takes the form: 
\begin{align}
	g_{\phi,0}(\tau)=\frac{C_\phi}{|\tau|^{2\delta}}~~~(\frac{1}{2}\leq \delta<1),  
\end{align}
where a convenient range for the $\phi$-field scaling dimension $\delta$ is imposed. Physically, $\delta = 1/2$. Equations \eqref{TensorRepQIEq} and \eqref{TensorRepREq} then imply that at $\tau^{-1}\ll \Lambda$, 
\begin{align}
	Q_A(\tau)&=J_0^2C_\phi A\frac{\sign(\tau)}{|\tau|^{2(\Delta+\delta)}},\\
	R(\tau)&=\rmi J_0^2C_\phi B\frac{1}{|\tau|^{2(\Delta+\delta)}}. 
\end{align}
In order to solve \eqref{TensorImpCompactEqGInv}, Green's functions in the frequency space are needed. Assuming $0<2\Delta<1$, we find that 
\begin{align}
	G_A(\rmi\omega)&=A\left[ 2\rmi\cos(\pi\Delta)\Gamma(1-2\Delta)\sign(\omega)|\omega|^{2\Delta-1} \right], \\
	G_{AB}(\rmi\omega)&=\rmi B\left[ 2\sin(\pi\Delta)\Gamma(1-2\Delta)|\omega|^{2\Delta-1} \right]. 
\end{align}
These can be obtained by directly Fourier transforming the power-law ansatz as if it was valid for all $\tau$. However, as the leading behavior at small $\omega$, the above results are more generally correct; with our assumptions for $\Delta$, both $G_A(\rmi\omega)$ and $G_{AB}(\rmi\omega)$ have power-law divergences at $\omega\rightarrow 0$, and the leading terms are completely determined by the slowly decaying long-time behavior of $G_A(\tau)$ and $G_{AB}(\tau)$. The self-energy functions $Q_I$ and $R$ are trickier to deal with. We assume in addition that $1< 2(\Delta+\delta)<2$, then the Fourier transform of $Q_A(\tau)$ can be directly obtained using the ansatz as well as principal value integral near $\omega=0$: 
\begin{widetext}
\begin{align}
	Q_A(\rmi\omega)=J_0^2C_\phi A\left[ 2\rmi\cos(\pi(\Delta+\delta))\Gamma(1-2(\Delta+\delta))\sign(\omega)|\omega|^{2(\Delta+\delta)-1} \right]. 
\end{align}
\end{widetext}
As $\omega\rightarrow 0$, the above expression has no divergence, thus one may wonder whether this converging power-law behavior is still only determined by the long-time limit of $Q_A(\tau)$. We note that $\pa Q_A(\rmi\omega)/\pa\omega$ does have a divergence at $\omega=0$ whose leading term is determined by the slowly decaying tail of $\tau Q_A(\tau)$. Therefore, \emph{modulo a constant term}, this expression is indeed the leading behavior of $Q_A(\rmi\omega)$ near $\omega=0$ which depends only on the long-time form of $Q_A(\tau)$. Moreover, the oddness of $Q_A(\rmi\omega)$ guarantees that the constant term $Q_A(\rmi\omega=0)$ actually vanishes. 

For the Fourier transform of $R(\tau)$, we assume 
\begin{widetext}
\begin{align}
	R(\rmi\omega=0)&=\rmi\xi,\\
	R(\rmi\omega)-R(\rmi\omega=0)&=\rmi J_0^2C_\phi B\int_{-\infty}^{\infty}\rmd\tau\left[ \frac{\rme^{\rmi\omega\tau}}{|\tau|^{2(\Delta+\delta)}}-\frac{1}{|\tau|^{2(\Delta+\delta)}} \right]\nonumber\\
	&=\rmi J_0^2C_\phi B\left[2\sin(\pi(\Delta+\delta))\Gamma(1-2(\Delta+\delta))|\omega|^{2(\Delta+\delta)-1} \right]. 
\end{align}
\end{widetext}
In the first integral expression of $R(\rmi\omega)-R(\rmi\omega=0)$, we did not mean that $R(\rmi\omega)$ is really given by $\rmi J_0^2C_\phi B|\tau|^{-2(\Delta+\delta)}$ for all $\tau$ (which does not have a Fourier transform); we are just using a convenient regularization. It is not hard to prove that, again, the final result only relies on $R(\tau\gg \Lambda^{-1})$. 
Plugging our ansatz into Eq.~\eqref{TensorImpCompactEqGInv}, we found that
\begin{align}
	&\Delta=\frac{1}{2}(1-\delta),\\
	&A^2\sin^2\left(\frac{\pi\delta}{2}\right)+B^2\cos^2\left(\frac{\pi\delta}{2}\right)=\frac{\delta\sin(\pi\delta)}{4\pi J_0^2C_\phi}. 
\end{align}
Notice that our previous assumptions on the scaling dimensions are met self-consistently. The ratio $A/B$ can only be determined using the fermion number constraint \eqref{FermionNumEq}. This is not an easy task; here we have only obtained the correlation functions in the long-time limit, but \eqref{FermionNumEq} is a UV equation. Therefore, we need to find a relation similar to the Luttinger theorem in a Fermi liquid which translates this UV constraint to an IR one, and this is done in Appendix~\ref{LuttingerThm}. Just quoting the result, we found
\begin{align}
	{B}/{A}={\tan(\theta_{0+})}/{\tan(\pi\Delta)}, 
\end{align}
where $\theta_{0+}$ is the unique solution within $[-\pi\Delta,\pi\Delta]$ to the equation
\begin{align}
\frac{\sin(2\theta_{0+})}{\sin(2\pi\Delta)}=\frac{1}{2\Delta-1}\left( 2\nu-1+\frac{2}{\pi}\theta_{0+} \right). 
\label{LuttingerThmEq}
\end{align}
The impurity spin operator is now $S_{\alpha\beta}=-\frac{1}{2}\rmi(\gamma_{A\alpha}\gamma_{A\beta}+\gamma_{B\alpha}\gamma_{B\beta})$ and its two-point function is $G_S(\tau)=\frac{1}{2}(A^2-B^2)/|\tau|^{4\Delta}$. 

\begin{figure}
	\centering
	\includegraphics[width=0.9\linewidth]{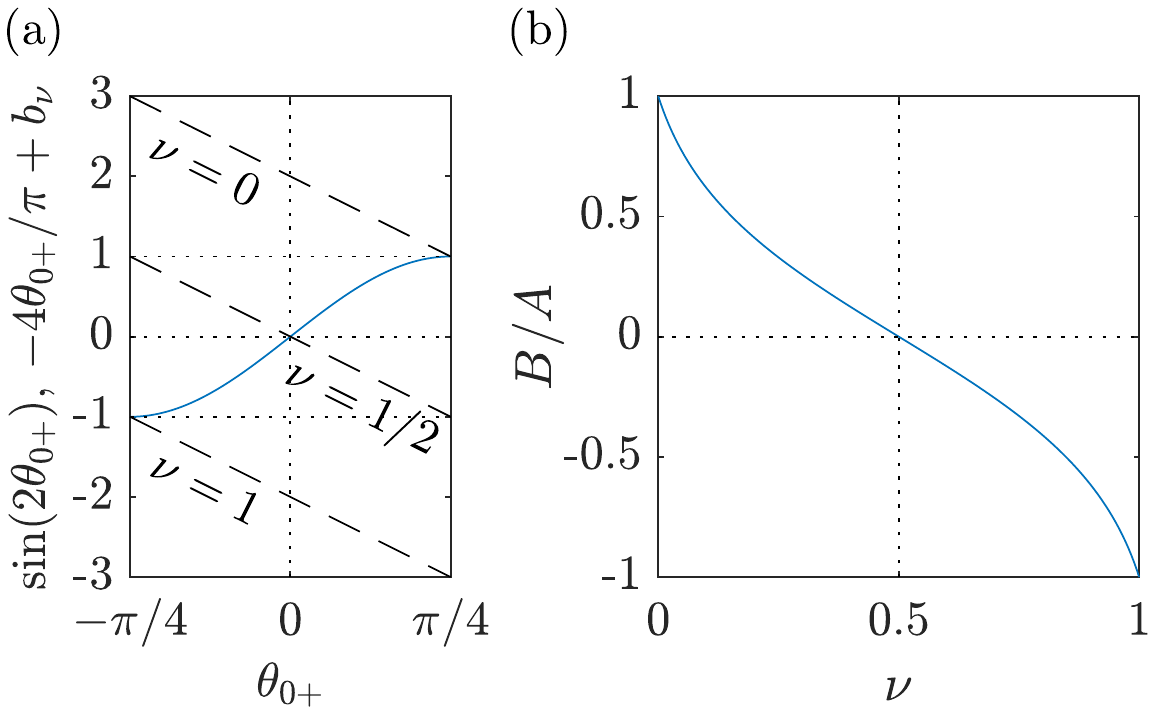}
	\caption{(a) Both sides of \eqref{LuttingerThmEqSpecial} as functions of $\theta_{0+}$. The blue solid line represents the left-hand side. The black dashed lines represent the right-hand side which is a linear function with intercept $b_\nu=-4\nu+2$, plotted for a few different values of $\nu$ as indicated in the figure. (b) Solution for $B/A$ as a function of $\nu$ in the case $\Delta=1/4$. }
	\label{LuttingerPlot}
\end{figure}

Plugging in the values $\delta=1/2$ and $C_\phi=1/(4\pi)$, we have $\Delta=1/4$, $A^2+B^2=1/J_0^2$ and $B/A=\tan\theta_{0+}$. Recalling $A\geq0$, these yield
\begin{align}
	A=\frac{\cos\theta_{0+}}{|J_0|},\quad B=\frac{\sin\theta_{0+}}{|J_0|}, 
\end{align}
and $\theta_{0+}$ is determined by \eqref{LuttingerThmEq} which simplifies to 
\begin{align}
\sin(2\theta_{0+})=-\frac{4}{\pi}\theta_{0+}-4\nu+2. 
\label{LuttingerThmEqSpecial}
\end{align}
We plot both sides of this equation as well as the solution for $B/A$ as a function of $\nu$ in \figref{LuttingerPlot}. The particle-hole symmetry $\nu \to 1-\nu$ is clear from the plot. 
The impurity spin correlator is 
\begin{align}
	G_S(\tau)=\frac{\cos(2\theta_{0+})}{2J_0^2|\tau|}. 
\end{align}

We conclude this section by noting that an alternative ansatz 
\begin{align}
	G_A=G_B=0,\quad G_{AB}(\tau)=\pm\frac{\sign(\tau)}{J_0\sqrt{|\tau|}}, 
\end{align}
which breaks the $U(1)$ gauge symmetry also solves the \emph{long time limit} of the self-consistent equations. However, we believe this is not a physical solution because it gives the same nonzero spin correlation function for all filling $\nu$, including $\nu=0$ where we expect the correlator to vanish. 

\subsection{Correlator at nonzero temperature and susceptibility}
In this subsection, we compute the susceptibility to a uniform external magnetic field as we did in the spinor cases. 

\begin{figure}%[h]
	\centering
	\includegraphics[width=0.7\linewidth]{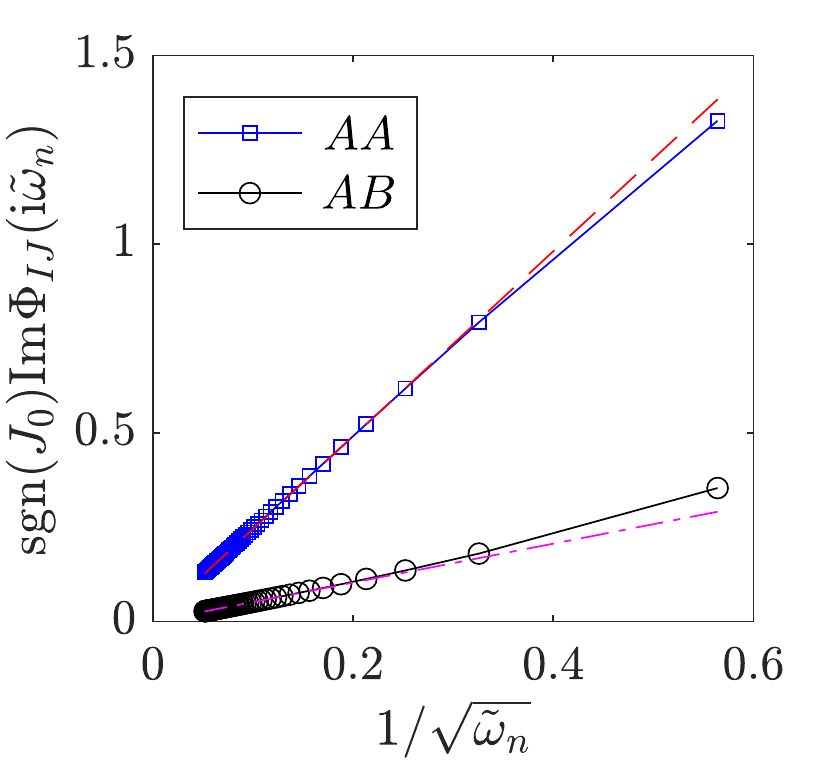}
	\caption{$\sign(J_0)\imag\Phi_{IJ}(\mathrm{i}\tilde\omega_n)$ as a function of $1/\sqrt{\tilde\omega_n}$ for $60$ positive values of $\tilde\omega_n$. Blue squares and black circles represent $IJ=AA$ and $IJ=AB$, respectively. The filling fraction is estimated to be $\nu=0.3354\pm0.0001$ by a linear fitting of $F_2(\rmi\tilde\omega_n)/F_1(\rmi\tilde\omega_n)$ with respect to $1/\tilde\omega_n$ using the $13$ smallest positive frequency values. 
	Red dashed lines and purple dotted dashed lines indicate the asymptotic (zero-temperature) behavior for $\nu=1/3$. }
	\label{TensorImpCorrelatorFiniteTem}
\end{figure}

After eliminating the $Q$ matrix from \eqref{TensorImpCompactEqGInv} and \eqref{TensorImpCompactEqQ}, the self-consistent equations become 
\begin{align}
	[G(\rmi\omega_n)]^{-1}=-\xi\sigma_y-\frac{J_0^2}{\beta}\sum_{\nu_n}g_{\phi,0}(\rmi\omega_n-\rmi\nu_n)G(\rmi\nu_n), 
\end{align}
where we have ignored the $-\rmi\omega_n/2$ term. 
Similar to the case of a spinor impurity coupled to tensor bulk fields, we define
\begin{align}
	J_0 G(\rmi\omega_n)=\frac{1}{\sqrt{T}}\Phi(\rmi\tilde\omega_n).  
\end{align}
$\Phi(\rmi\tilde\omega_n)$ takes the form $\Phi_A(\rmi\tilde\omega_n)\mathbbm{1}+\Phi_{AB}(\rmi\tilde\omega_n)\rmi\sigma_y$ and is purely imaginary. We can then rewrite the equation as 
\begin{align}
	[\imag\Phi(\rmi\tilde\omega_n)]^{-1}&=-\frac{\xi}{J_0\sqrt{T}}\rmi\sigma_y\nonumber\\&+\sum_{\nu_n}g_{\phi,0}(\rmi\omega_n-\rmi\nu_n)\imag\Phi(\rmi\tilde\nu_n). 
\end{align}
Using an isomorphism of (algebraic) fields: 
\begin{align}
	\{ a\mathbbm{1}+b(\rmi\sigma_y)|a,b\in\mathbb{R} \}&\cong \mathbb{C}. \\
	a\mathbbm{1}+b(\rmi\sigma_y)&\mapsto a+b\rmi\nonumber
\end{align}
the $2\times 2$ matrix function $\imag\Phi(\rmi\tilde\omega_n)$ can be conveniently represented by a complex number function $F=\imag\Phi_A+\rmi\imag\Phi_{AB}\equiv F_1+\rmi F_2$, satisfying
\begin{align}
F^{-1}(\rmi\tilde\omega_n)=-\frac{\rmi\xi}{J_0\sqrt{T}}+\sum_{\nu_n}g_{\phi,0}(\rmi\omega_n-\rmi\nu_n)F(\rmi\tilde\nu_n). 
\end{align}
$F_1$ is antisymmetric, thus the real part of the above equation becomes  
\begin{align}
	\frac{F_1(\rmi\tilde\omega_n)}{|F(\rmi\tilde\omega_n)|^2}=\frac{1}{4\pi}\sum_{\tilde\nu_n>0}\ln\left[ \frac{(\tilde\omega_n+\tilde\nu_n)^2+\mu^2}{(\tilde\omega_n-\tilde\nu_n)^2+\mu^2} \right]F_1(\rmi\tilde\nu_n). 
\end{align}
For the imaginary part of the equation, we eliminate $\xi$ from a specific frequency $\tilde\omega_n=\tilde\lambda$, leading to 
\begin{align}
	&-\frac{F_2(\rmi\tilde\omega_n)}{|F(\rmi\tilde\omega_n)|^2}+\frac{F_2(\rmi\tilde\lambda)}{|F(\rmi\tilde\lambda)|^2}+\frac{1}{4\pi}\sum_{\tilde\nu_n>0}\left\{ \ln\left[ \frac{(\tilde\omega_n-\tilde\nu_n)^2+\mu^2}{(\tilde\lambda-\tilde\nu_n)^2+\mu^2} \right]\right.\nonumber\\&\left.+
	\ln\left[ \frac{(\tilde\omega_n+\tilde\nu_n)^2+\mu^2}{(\tilde\lambda+\tilde\nu_n)^2+\mu^2} \right] \right\}F_2(\rmi\tilde\nu_n)=0. 
\end{align}
For example, one may take $\tilde\lambda=\pi$. We still need one more equation to effectively implement the fermion number constraint. In a numerical calculation where the frequency space is truncated, this is also necessary for matching the number of unknowns. From the zero temperature limit, we know that as $n\rightarrow\infty$, $F_2(\rmi\tilde\omega_n)/F_1(\rmi\tilde\omega_n)$ should approach the $B/A$ ratio. However, this is hard to impose in practice as we are not able to directly access the $n\rightarrow\infty$ limit. In our calculation, we constrained the ratio $F_2(\rmi\tilde\omega_n)/F_1(\rmi\tilde\omega_n)$ for one certain $\tilde\omega_n$, and then determined $F_2(\rmi\infty)/F_1(\rmi\infty)$ by an extrapolation procedure, which then gave an estimate of the filling fraction $\nu$. As before, when computing $F(\rmi\tilde\omega_n)$ up to some $\tilde\omega_{\rm max}$, we truncated the frequency space at a larger $\tilde\omega_{\rm ext}$ with $n_{\rm ext}=n_{\rm max}^2$. In \figref{TensorImpCorrelatorFiniteTem}, we present an example of the numerical solution to $F(\rmi\tilde\omega_n)$ where $\nu$ is closed to $1/3$. 

\begin{figure}
	\centering
	\includegraphics[width=1\linewidth]{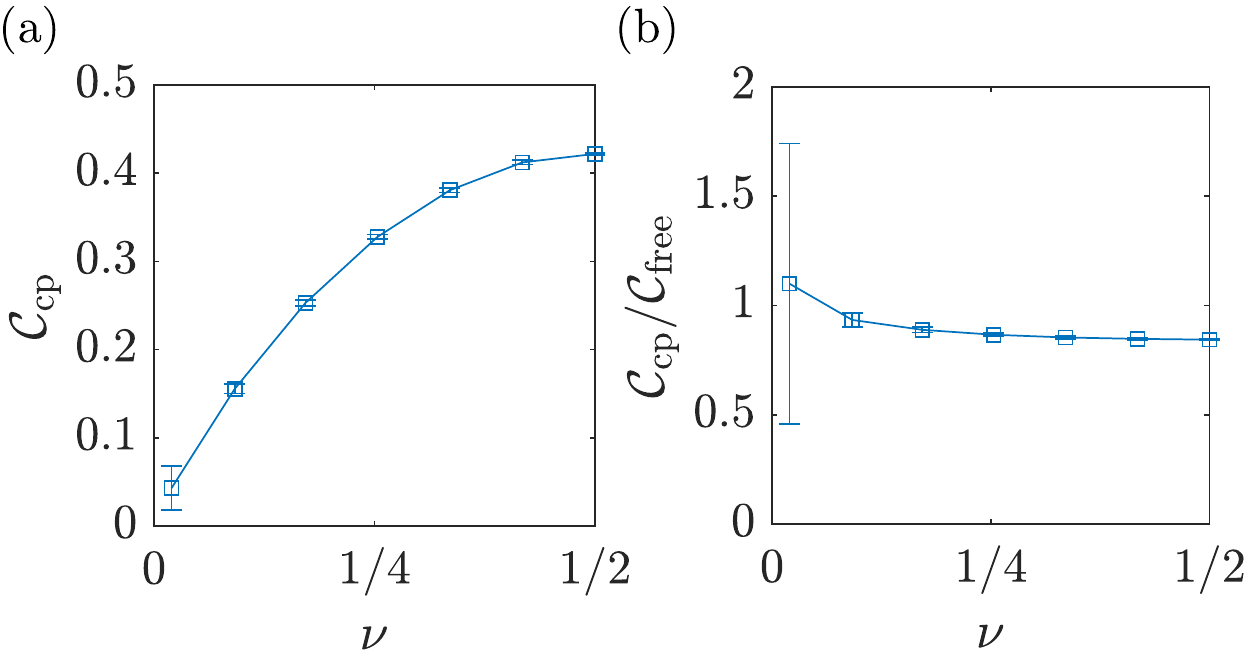}
	\caption{(a) $\mc C_{\rm cp}$ for an antisymmetric tensor impurity as a function of $\nu$ for $\nu\leq 1/2$. Error bars are indicated. Note that $\chi_{\rm imp}$ for the filling $\nu$ is the same as that for the filling $1-\nu$ due to the particle-hole symmetry. (b) $\mc C_{\rm cp}/\mc C_{\rm free}$ as a function of $\nu$. }
	\label{TensorImpSusceptibility}
\end{figure}

Computation of the susceptibility parallels the previous case. When the impurity is decoupled, we found
\begin{align}
	\text{(free imp.)}\quad\chi_{\rm imp}=\frac{2\nu(1-\nu)N}{N-1}\frac{1}{T}
	\stackrel{N\rightarrow\infty}{\longrightarrow}2\nu(1-\nu)/T. 
\end{align}
For example, when $N=3,~\nu=1/3$, the result coincides with $S(S+1)/(3T)$ for $S=1$, as expected. 
After turning on the coupling, with the same Feynman diagrams as in \figref{SusceptibilityDiagrams}, we find $\chi_{\rm imp}={\mc C}_{\rm cp}/T$. Again there is no anomalous dimension in $T$. Our numerical result for $\mc C_{\rm cp}$ is shown in \figref{TensorImpSusceptibility}. 
The computation strategy is similar to that in Section~\ref{SpinorImpTensorBulk}. For the special filing $\nu=1/2$, $F_2=0$ due to the particle-hole symmetry and we simply multiplied the result in the spinor case by a factor of $2$. For other values of $\nu$, we solved $F(\rmi\tilde\omega_n)$ for $n_{\rm max}=60$, extended the solution using the known asymptotic form, and then obtained an extrapolated result for the susceptibility at infinite frequency cutoff. We collected three sources of error: one from the extrapolation fitting for the susceptibility, another one from varying $n_{\rm max}$ and a new one from the estimation of $\nu$ (another extrapolation) as explained earlier. 

From our result for the susceptibility, at least at this particular fixed point, the antisymmetric tensor impurity is not fully screened by the bulk. This may sound surprising because the impurity now transforms under a regular representation of $SO(N)$. In fact, there is still a secret symmetry fractionalization happening. Suppose $N$ is even, the $\mathbb{Z}_2$ center of $SO(N)$ (generated by $-\mathbbm{1}_{N\times N}$) has a trivial action on the bulk fields $\phi_{\alpha\beta}$, i.e. the low-energy symmetry in the bulk is truly $SO(N)/\mathbb{Z}_2$. In contrast, this $\mathbb{Z}_2$ center symmetry acts nontrivially on the impurity Hilbert space when $\nu N$ is odd. This argument does not apply to odd $N$ or even $\nu N$, but for large $N$ the continuity of our solution in $N$ and $\nu$ implies the existence of a fixed point with only partial screening in these cases also. In Section~\ref{TensorImpVecBulk}, we will consider an impurity problem free of similar symmetry fractionalization and there we, indeed, find a fully screened fixed point. 

\section{A $\mathbb{Z}_2$ symmetric impurity in the $O(3)$ model}\label{TensorImpVecBulk}
\begin{figure}
    \centering
    \includegraphics{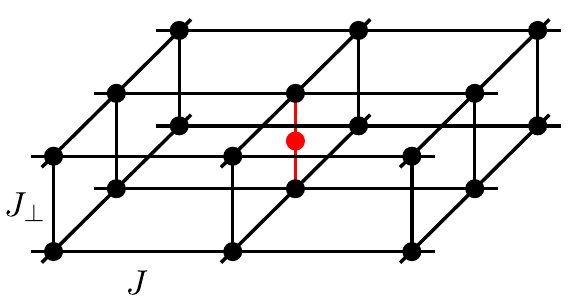}
    \caption{Bilayer antiferromagnet and an impurity (marked as a red dot) symmetric under the $\mathbb{Z}_2$ layer interchange symmetry. }
    \label{BilayerAFM}
\end{figure}
As our final model, we come back to a spin $S$ impurity in the $O(3)$ model, Eq.~(\ref{O3Sb}). However, we now require that the impurity respect the unitary $\phi_\alpha \to -\phi_\alpha$ symmetry. For instance, we can consider the $O(3)$ transition in a bilayer antiferromagnet.\cite{SandvikBL} The layer exchange symmetry acts as $\phi_\alpha \to -\phi_\alpha$. We place the impurity midway along a rung of the bilayer and require it to couple in a way that respects the layer exchange symmetry. See Fig.~\ref{BilayerAFM} for an illustration. In this setup the coupling $L_{\rm cp} \sim S_\alpha \phi_\alpha$ of the impurity spin operator $S_\alpha$ to the bulk fields is prohibited. The next most relevant coupling is
\begin{align} S_{\rm cp} = -J \int d \tau \, O_{\alpha \beta}(\tau) t_{\alpha \beta}(\tau, \bs r=0) \label{ST} \end{align}
where
\begin{align} O_{\alpha \beta} &= \frac{1}{2} (S_\alpha S_\beta + S_\beta S_\alpha) - \frac{\delta_{\alpha \beta}}{3} S(S+1) \label{Odef}\\
t_{\alpha \beta} &\sim \phi_\alpha \phi_\beta - \frac{1}{3} \delta_{\alpha \beta} (\vec{\phi}^2)  \end{align}
We note that $O_{\alpha \beta} = 0$ for spin $S=1/2$, thus, we only consider $S \ge 1$ below.

The traceless symmetric tensor $t_{\alpha\beta}$ operator in the $O(3)$ model is known to have a scaling dimension $\Delta_t = 1.20954(32)$.\cite{O3bootstrap} Thus, the impurity coupling $J$ in Eq.~(\ref{ST}) is slightly irrelevant, $d J/d\ell = -(\Delta_t - 1) J$. (Here and below $x \sim e^\ell$). This means that for small enough $J$ the impurity decouples from the bulk, forming an essentially free spin. 

We next perform a perturbative renormalization group analysis in $J$ searching for non-trivial weak coupling fixed points, using  $\Delta_t - 1 \approx 0.21$ as an expansion parameter. We note that the next most relevant impurity coupling is $\delta S \sim  \lambda \int d\tau s(\tau, 0)$, where $s(x)$ is the leading $SO(3)$ scalar. The scaling dimension $\Delta_s = 1.59488(81)$ (see \cite{O3bootstrap}), thus, this coupling is somewhat more irrelevant than $J$ and will not be included in our analysis.

Applying standard perturbation theory in $J$ to second order we obtain (see appendix \ref{app:PertTensor}):
\begin{align}
    \frac{d J}{d\ell} = -(\Delta_t - 1) J + a J^2 \label{betaJ}
\end{align}
where the coefficient $a$ is given by
%\begin{align} a = \frac{4}{3} \left(S(S+1) - \frac{15}{4}\right) \lambda_{ttt}\end{align}
\begin{align} a = \frac{\sqrt{2}}{3 } \left(S(S+1) - \frac{15}{4}\right) \lambda_{ttt}\end{align}
and $\lambda_{ttt} \approx 1.499$ is the  coefficient of $t$ in the $t\times t$ OPE (see appendix \ref{app:PertTensor} for our normalization convention\footnote{We thank Shai Chester for explaining the normalization conventions in Ref.~\onlinecite{O3bootstrap} to us}).\cite{O3bootstrap}  Intriguingly, $a < 0$ for $S = 1$, $a = 0$ for $S = 3/2$, and $a > 0$ for $S > 3/2$. This means that for $S = 1$ we have an IR unstable fixed point at ${\it negative}$ $J_c = - (\Delta_t - 1)/|a|$, and for $J < -|J_c|$, $J$ runs away to $-\infty$. As we discuss below, we expect this runaway flow is towards a fully screened impurity phase. On the other hand, for $S \ge 2$ we have an unstable  fixed point at positive $J_c = (\Delta_t - 1)/a$, and for $J > J_c$ $J$ runs away to $+ \infty$. We do not currently know the nature of the $J = +\infty$ fixed point, but we don't expect the impurity to be fully screened here. (Indeed, this flow occurs both for integer and half-integer $S \ge 2$, and for half-integer $S$ we don't expect full screening to occur.) Finally, for $S = 3/2$, $a = 0$ and any perturbatively accessible fixed points will be controlled by the $O(J^3)$ term in $dJ/dl$. Below, we exclude $S = 3/2$ from our discussion. 

From the $\beta$-function (\ref{betaJ}) we conclude that the correlation length exponent $\nu$  at the critical point $J_c$ ($\xi \sim |J- J_c|^{-\nu}$) is given by
\begin{align} \nu^{-1} = \Delta_t - 1 \approx 0.21 \end{align}
both for $S = 1$ and for $S \ge 2$. We also compute perturbatively the scaling dimension of the impurity spin $S^a$ at the critical point $J = J_c$ (see appendix \ref{app:PertTensor}):
%\begin{align}  \Delta_{S^a} &\approx 2 (S (S+1) - 3/4) J^2 \nonumber \\ &\approx \frac{9 (\Delta_t - 1)^2}{8 \lambda_{ttt}^2} \frac{S(S+1) - 3/4}{(S(S+1) - 15/4)^2} \end{align}
\begin{align}  \Delta_{S^a} &\approx 2 (S (S+1) - 3/4) J^2 \nonumber \\ &\approx \frac{9 (\Delta_t - 1)^2}{ \lambda_{ttt}^2} \frac{S(S+1) - 3/4}{(S(S+1) - 15/4)^2}  \nonumber\\ &\approx  0.18\frac{S(S+1) - 3/4}{(S(S+1) - 15/4)^2}  \label{DeltaST} \end{align}

We now say a few words about the nature of the $J = -\infty$ phase that appears to be realized for $S =1$ and $J < -|J_c|$. We believe this is a fully screened phase. There are several indications that this is the case. Most heuristically, let's begin by freezing the direction of the bulk order parameter $\phi_\alpha$ (e.g. along the $z$ direction). The impurity coupling (\ref{ST}) is then $\delta H \sim - J S^2_z$. For $J < 0$ this coupling favors the state with minimal $S_z$. For integer $S$, this is the unique state $S_z = 0$. Now, turning back on slow fluctuations of $\phi_\alpha$, the impurity spin will be in the instantaneous eigenstate of $\phi_\alpha S_\alpha$ with zero eigenvalue, which produces no Berry phase. This absence of Berry phase makes us conclude that for $J \to -\infty$ the magnetic impurity essentially acts as a non-magnetic one - i.e. full screening occurs.

Another indication that the $J = -\infty$ fixed point for $S = 1$ is fully screened comes from considering the large $N$ generalization of the current problem that we discuss below. This generalization is quite similar to the $SU(N)$ impurity model studied in Ref.~\cite{FlorensPRL}.

\subsection{Large $N$ generalization}

We consider a vector $O(N)$ model in the bulk, Eq.~(\ref{VectorBulkAction}). The impurity is taken to transform in the traceless symmetric tensor representation of $SO(N)$ with $N_b$ indices. There is a unique traceless symmetric two-index tensor operator $O_{\alpha \beta}$  acting on the impurity Hilbert space, 
which can be taken as
\begin{align} 
O_{\alpha \beta} = -\frac{1}{2}(S_{\alpha \gamma} S_{\beta \gamma} + S_{\beta \gamma} S_{\alpha \gamma}) + \frac{2}{N} \delta_{\alpha \beta} C
\label{ODefGeneralN}
\end{align}
with $S_{\alpha \beta}$ the generators of $SO(N)$ in the representation considered and $C = \frac{1}{2} S_{\alpha \beta} S_{\alpha \beta} =   N_b (N_b + N -2)$ - the quadratic Casimir.\footnote{We've normalized $O_{\alpha \beta}$ so that it reduces to (\ref{Odef}) for $N = 3$.} We consider the impurity coupling:
\begin{align} 
S_{\rm cp} = - J \int \, d\tau O_{\alpha \beta}(\tau) t_{\alpha \beta}(\tau,\bs r=0) \label{OtN} 
\end{align}
with $t_{\alpha \beta}$ - the bulk symmetric tensor
\begin{align} 
t_{\alpha \beta} \sim \phi_\alpha \phi_\beta - \frac{\delta_{\alpha \beta}}{N} (\vec{\phi}^2)
\label{tDefGeneralN}
\end{align}
The resulting model is similar to the large-$N$ model considered in Ref.~\onlinecite{FlorensPRL}, except the symmetry group is $SO(N)$ instead of $SU(N)$. At large $N$ the scaling dimension $\Delta_t$ is given by\cite{GraceyON}
\begin{align} \Delta_t = 1 + \frac{32}{3 \pi^2 N} + O(1/N^2)  \end{align}
Thus, the impurity coupling $J$ is slightly irrelevant in the large-$N$ limit, so the impurity forms a free spin for sufficiently small $J$. 

We again perform a perturbative RG analysis in $J$ utilizing the parametric smallness of $\Delta_t - 1$ (see appendix \ref{app:PertTensor}). We obtain a flow equation as in (\ref{betaJ}) but with the coefficient $a$ now given by
\begin{align} a &= - \frac{\sqrt{2}\lambda_{ttt}}{4 N}  (4 (N-4) N^2_b + 4 (N^2 - 6N + 8) N_b \nonumber\\ &+ N (N^2-4N+8))\end{align}
where the OPE coefficient $\lambda_{ttt} \to 2$ for $N \to \infty$. Note that $a <0$ for all integer $N > 3$ independent of $N_b$. Thus, we expect that the critical point with $J_c > 0$ found for $S \ge 2$ in the $N = 3$ case does not exist for $N \ge 4$ (or at least is not perturbatively acessible). Instead, for $N \ge 4$, the critical coupling $J_c = (\Delta_t -1)/a$ is negative, and for $J < -|J_c|$ we have a runaway flow to $J = -\infty$ as in the $S = 1$, $N = 3$ case. In particular, taking the limit $N \to \infty$ with $\nu = N_b/N$-fixed, we have 
\begin{align} a = - 2 \sqrt{2} N^2 (\nu + 1/2)^2, J_c =   -\frac{8 \sqrt{2}}{3 \pi^2 N^3 (\nu + 1/2)^2} \label{JcN} \end{align}
One may ask whether perturbation theory in $J$ continues to work for the coupling $J_c \sim O(1/N^3)$. As we explain below, the answer is yes. 

\subsubsection{Beyond perturbation theory}
\label{TensorSymmSaddle}

Next, to understand the nature of the flow $J \to -\infty$ for $J < J_c$ we set-up a non-perturbative large-$N$ calculation. We can represent our impurity in terms of canonical complex boson modes $b_\alpha$. The constraint of $N_b$ boxes in the symmetric $SO(N)$ tensor representation translates to $b^{\dagger}_\alpha b_\alpha |\psi\rangle = N_b |\psi\rangle$, where $|\psi\rangle$ is a state in the physical subspace $V_{phys}$. The traceless condition translates to $b_\alpha b_\alpha |\psi\rangle = 0$. The action of the operator $O_{\alpha \beta}$ on $|\psi\rangle \in V_{phys}$ is
\begin{align}
     O_{\alpha \beta} &\to -\frac12 (N+2 N_b - 4) \left(b^{\dagger}_{\alpha} b_{\beta} + b^{\dagger}_\beta b_\alpha -2 \frac{N_b \delta_{\alpha \beta}}{N}\right) \nonumber\\ &+ (b^{\dagger}_\gamma b^{\dagger}_{\gamma}) b_{\alpha} b_{\beta} \label{Ob}
\end{align}
In particular, when calculating matrix elements of $O_{\alpha \beta}$ within  $V_{phys}$ we may drop the last term in Eq.~(\ref{Ob}). Thus, we consider the action
\begin{align}
	S_{\rm imp}&=\int\rmd\tau b^\da_\alpha\pa_\tau b_\alpha+\int\rmd\tau~\xi(\tau)\left( b^\da_\alpha b_\alpha-\nu N \right) \nonumber\\
	&+ V \int d\tau  b^{\dagger}_\alpha b^{\dagger}_\alpha b_{\beta} b_{\beta}, \label{V}\\
	S_{\rm cp}&= \frac{2 \pi \sqrt{2} J_0}{N} \int\rmd\tau b^{\dagger}_\alpha b_\beta \phi_\alpha\phi_\beta (\bs r=0). \label{J0} 
\end{align}
Here $\xi(\tau)$ is a Lagrange multiplier enforcing $b^{\dagger}_\alpha b_\alpha |\psi\rangle = N_b |\psi\rangle$. We enforce the constraint $b_{\alpha} b_{\alpha} |\psi\rangle = 0$ energetically by sending the energy $V \to \infty$. The coupling $J_0$ is related to $J$ in Eq.~(\ref{OtN}) via $J_0 = N (N+2 N_b - 4)J$ and the factor $2\pi \sqrt{2}$ in (\ref{J0}) is introduced to match the normalization of $t_{\alpha \beta}$ in the large-$N$ limit (see appendix \ref{app:PertTensor}). 

We can decouple the quartic terms in (\ref{V}), (\ref{J0}):
\begin{align} S_{\rm imp}&=\int\rmd\tau b^\da_\alpha\pa_\tau b_\alpha+\int\rmd\tau~\xi(\tau)\left( b^\da_\alpha b_\alpha-\nu N \right) \nonumber\\
&+ i \int d\tau\, \left(u b^{\dagger}_{\alpha} b^{\dagger}_{\alpha} + u^* b_{\alpha} b_{\alpha}\right) + \frac{1}{V} \int d \tau u^* u  \label{Su} \\
S_{\rm cp} &= \int d \tau \left(s b^{\dagger}_\alpha \phi_\alpha(\bs r=0) + s^* b_\alpha \phi_\alpha(\bs r=0)\right) \nonumber\\ &-\frac{N}{2 \pi \sqrt{2} J_0} \int d \tau s^* s  
\end{align}
$s(\tau)$ and $u(\tau)$ are auxiliary fields. The limit $V \to \infty$ can now be taken by simply dropping the last term in (\ref{Su}). 
It is now clear that the theory possesses a large-$N$ limit with fixed $J_0$, $\nu$ where fluctuations of $\xi$, $s$ and $u$ are suppressed. Note that taking $J_0 \ll 1$ only suppresses fluctuations of $s$ further - in this regime we can perform combined expansion in $J_0$ and $1/N$, as we have done in the previous section ($J_0 \ll 1$ translates to $J \ll 1/N^2$ and our $J_c$ in Eq.~(\ref{JcN}) lies in this range). Below, we focus on $J_0 \sim O(N^0)$ in order to understand the phase diagram of our model.

As already noted, when $N = \infty$ the fields $\xi$, $s$, $u$ (and $\lambda(x)$) become frozen at their saddle-point value. The saddle-point equations are:
\begin{align} &\langle b^{\dagger}_\alpha b_{\alpha} \rangle = \nu N, \quad \langle b_\alpha b_\alpha \rangle = 0\\
& \langle b_\alpha \phi_\alpha(\bs r=0) \rangle = \frac{N}{2 \pi \sqrt{2} J_0} s, \quad \langle b^{\dagger}_\alpha \phi_\alpha(\bs r=0) \rangle = \frac{N}{2 \pi \sqrt{2} J_0} s^*\\
&\langle \phi_\alpha\phi_\alpha(x) \rangle = \frac{N}{g} \label{saddleb} \end{align}

One saddle-point is $s = s^* = u = u^* = 0$. Here the impurity and the bulk decouple. Working at finite temperature $T$ we  then find the familiar saddle-point $\lambda(x) = m^2 = \mu T$ for the bulk, whereas $\xi(\tau) = T \log(\nu^{-1}+1)$, so that the $b$ propagator $G_{b}(\tau) = \langle b_\alpha(\tau) b^{\dagger}_\alpha (0)\rangle= \frac{1}{\beta} \sum_{\omega_n} \frac{1}{-i \omega_n + \xi}$. It is instructive to compute the $s$ propagator at this decoupled fixed point. Letting $G_{s}(i \omega_n) = \int d\tau \langle s(\tau) s^*(0)\rangle e^{i \omega_n \tau}$, we obtain 
\begin{align} G_s(i \omega_n) &= \frac{1}{N}\left(-(2 \pi \sqrt{2} J_0)^{-1} + \Pi_s(i \omega_n)\right)^{-1}\\
\Pi_s(i \omega_n) &= - \int d \tau G_b(\tau) g_{\phi,0}(\tau) e^{i \omega_n \tau}\end{align}
with $g_{\phi,0}$ given by Eq.~(\ref{gphi0}). From the small $\tau$ behavior of $G_b(\tau)$, $g_{\phi,0}(\tau)$, the UV divergent part of $\Pi_s(i \omega_n) \stackrel{UV}{=} -\frac{1}{4\pi} (2 \nu +1) \log (\Lambda/|\omega_n|)$, leading to the RG flow
\begin{align} \frac{d J_0}{d\ell} = - \sqrt{2} (\nu+\tfrac12) J^2_0 \end{align}
which agrees with our perturbative result in Eqs.~(\ref{betaJ}), (\ref{JcN}). Thus, at $N = \infty$, $J_0 > 0$ runs logarithmically to zero and $J_0 < 0$ gives a run-away flow $J_0 \to -\infty$. We also now have a hint of the nature of this run-away flow. For $J_0 < 0$ we have 
\begin{align} G_s(i\omega_n = 0)^{-1} =  \frac{N}{2 \pi \sqrt{2}} \left(|J_0|^{-1} - \sqrt{2} (\nu+\tfrac12) \log(\Lambda/T)\right) \end{align}  
Thus, $G_s^{-1}(i \omega_n = 0)$ switches sign and becomes negative for  $T < T_0= \Lambda \exp\left[-\left((\nu+\tfrac12)\sqrt{2} |J_0|\right)^{-1} \right]$. This implies that the $s = 0$ saddle-point becomes unstable for $T < T_0$ and suggests that the true saddle-point in this regime should have $\langle s \rangle \neq 0$. (Note that once $1/N$ corrections are included, we expect no true phase transition at $T_0$, but rather a crossover.)

Thus, we expect the $T = 0$ saddle point for $J_0 < 0$ has $s \sim \langle b_{\alpha} \phi_\alpha \rangle \neq 0$ (and also by symmetry $u \neq 0$ and $u^* \neq 0$). We will not attempt to solve the saddle-point equations (\ref{saddleb}) in this regime. (At such a saddle $\lambda(x)$ acquires spatial dependence and has to be solved for self-consistently.) We expect that this saddle-point describes a fully screened impurity: the $b$ propagator will be non-singular near $\omega = 0$, so that the impurity modes can be safely integrated out. 

\section{Comparison with Related Results}\label{ComparisonSection}
\subsection{Impurity problems at $N=3$}\label{DifferentApproaches}
It is interesting to compare our $1/N$-expansion results with the results of different approaches to related problems. The  $\epsilon$-expansion with $\epsilon=4-D$ \cite{SubirImp} and quantum Monte Carlo (QMC) studies \cite{SandvikScalingDim,SandvikCurie} have been applied to the $SO(3)$ (or $Spin(3)$) symmetric impurity problem where a spin-$S$ impurity is coupled to the 2+1D $O(3)$ critical bulk. 

First consider the case of a spin-$1/2$ impurity for which the full action is given in Eqs.~\ref{O3TotalAction}-\ref{O3Scp}. 
%at the beginning of Section~\ref{SO3Problem}. 
Model 1 (Section~\ref{SpinorImpVecBulk}, Eqs.~\ref{Model1TotalAction}-\ref{Model1Scp}) and Model 2A (Section~\ref{SpinorImpTensorBulk}, Eqs.~\ref{TensorBulkAction} and \ref{Model2ASimpScp}) are both $SO(N)$ generalizations of it. 
In Table~\ref{Table:Comparison}, we list the results from different approaches on (1) the scaling dimension of the spin operator $S_\alpha$ or its $SO(N)$ analogs and (2) ${\mc C}_{\rm cp}/{\mc C}_{\rm free}$, the ratio between the renormalized and free Curie coefficients. See the table caption for more details such as the expansion orders. Notice that there are two inequivalent generalizations of the spin operators to $N>3$, one proportional to $\rmi\gamma_0\gamma_\alpha$ and the other proportional to $\rmi\gamma_\alpha\gamma_\beta$, appearing in both Model 1 and 2A. There is no natural way to prefer one or the other in comparing to $N = 3$ results, thus the scaling dimensions for both sets of operators are given in the table, denoted as $\Delta_{\rm vec}$ and $\Delta_{\rm adj}$, respectively. Let us also quote the analytic formulas from the $\epsilon$-expansion: 
\begin{align}
	&[S_\alpha]=\frac{1}{2}\epsilon-\epsilon^2\left( \frac{5}{484}+\frac{\pi^2}{11}[S(S+1)-1/3] \right)+\cdots\label{epsilonExpScalingDim}\\
	&{\mc C}_{\rm cp}/{\mc C}_{\rm free}=1+\sqrt{\frac{33\epsilon}{40}}-\frac{7\epsilon}{4}+\cdots \label{epsilonExpCurie}
\end{align}
Note that ${\mc C}_{\rm cp}/{\mc C}_{\rm free}$ is $S$-independent to this order in $\epsilon$. We emphasize that when using the above expressions to obtain concrete numbers, we have directly plugged $\epsilon=1$ and the value of $S$ into the power series; usually better results can be obtained using Pad\'{e} approximants or other resummation techniques. In both large-$N$ models, $\Delta_{\rm vec}$ and $\Delta_{\rm adj}$ differ a lot from each other, which indicates a sensitive dependence on $N$ of at least one of the two scaling dimensions. Nonetheless, $\Delta_{\rm vec}$ of Model 2A does roughly agrees with the QMC result. Also, both large-$N$ models predict that the impurity operator which directly couples to the bulk field has a scaling dimension $1/2$, coinciding with the 1-loop $\epsilon$-expansion. 
Our predictions for the Curie coefficient are numerically  much closer  to the QMC finding of ${\mc C}_{\rm cp}/{\mc C}_{\rm free} \sim 1.04$ than the $\epsilon$-expansion, but there is still a qualitative difference: in QMC  the impurity susceptibility is slightly  {\em enhanced} by the bulk instead of being suppressed. 
\begin{table}
	\centering
	\begin{tabular}{|c|c|c|c|}
		\hline
		{\bf Methods} & $\boldsymbol{\Delta_{\rm vec}}$ & $\boldsymbol{\Delta_{\rm adj}}$ & $\boldsymbol{{\mc C}_{\rm cp}/{\mc C}_{\rm free}}$  \\
		\hline
		
		\makecell{Model 1} & $1/2$ & $0$ & $0.860$ \\
		%\hline
		
		\makecell{Model 2A} & $1/4$ & $1/2$ & $0.844(7)$ \\
		%\hline
		
		\makecell{$\epsilon$-Exp.} & $0.116$ & $\leftarrow$ & $0.158$ \\
		%\hline
		
		\makecell{QMC} & $0.20(1)$ & $\leftarrow$ & $1.048(8)$ \\
		\hline
		
	\end{tabular}
	\caption{Comparison between different approaches for the problem of a spin-$1/2$ impurity coupled to the 2+1D $O(3)$ critical bulk. For the two large-$N$ models (Model 1 and 2A) considered in this paper, we display both the scaling dimension $\Delta_{\rm vec}$ of $S_\alpha\propto\rmi\gamma_0\gamma_\alpha$ that transforms as an $SO(N)$ vector and the scaling dimension $\Delta_{\rm adj}$ of $S_{\alpha\beta}\propto\rmi\gamma_\alpha\gamma_\beta$ that transforms in the adjoint representation. For the actual $SO(3)$ problem ($N=3$) studied previously by both the $\epsilon$-expansion and QMC methods, these two set of operators are equivalent. The last column displays ${\mc C}_{\rm cp}/{\mc C}_{\rm free}$. The $\epsilon$-expansion results are of 2-loop order. The numerical error for ${\mc C}_{\rm cp}/{\mc C}_{\rm free}$ of Model 1 is not shown since it is smaller than the last digit retained here. }
	\label{Table:Comparison}
\end{table}

Next consider the case of a spin-$1$ impurity where less data is available. Model 2B (Section~\ref{TensorImpTensorBulk}) with either $\nu=1/N$ or $\nu=1/3$ is an $SO(N)$ generalization of this problem. The $\epsilon$-expansion gives $[S_\alpha]=1/2$ at 1-loop order and a negative number at 2-loop order. Once again, our prediction $[S_{\alpha\beta}]=1/2$ matches the 1-loop result. We are not aware of any QMC study of this scaling dimension. Regarding the susceptibility, 2-loop $\epsilon$-expansion gives the same ratio ${\mc C}_{\rm cp}/{\mc C}_{\rm free}=0.158$ and QMC gives ${\mc C}_{\rm cp}/{\mc C}_{\rm free}=0.995(3)$. These should be compared with Fig.~\ref{TensorImpSusceptibility}b either at $\nu=1/3$ or in the $\nu\rightarrow 0$ limit. It seems that the large-$N$ value of the susceptibility ratio ${\mc C}_{\rm cp}/{\mc C}_{\rm free}$ is closed to one for all $\nu$ and thus compatible with the QMC prediction (although we are less certain about the $\nu\rightarrow 0$ limit due to the huge error bar). 

\subsection{Sachdev-Ye-Kitaev models}\label{SYKConnections}
The SYK models \cite{Kit.KITP.1,Kit.KITP.2,SYKRemarks,SubirCplxSYK,GuCplxSYK} are 0+1D quantum models which have been extensively studied in the large-$N$ limit, that exhibit a conformal structure. We can compare correlators in the SYK  models with  correlators of our 0+1D quantum impurity, which is immersed in a 2+1D conformal bulk. 

Models 2A (Section~\ref{SpinorImpTensorBulk}) and 2B (Section~\ref{TensorImpTensorBulk}) have strong similarities to the real (Majorana) and complex SYK models, respectively, at least at zero temperature. We will elaborate on this connection in this subsection. For our purpose here, we will regard the fermion operators in these two models as physical\footnote{We were in fact already doing so when looking for large-$N$ solutions to these two models earlier.}, i.e. we do not apply the fermion-number parity projection in Model 2A and treat $\xi$ as a real constant in Model 2B instead of a Lagrange multiplier. 

{\bf Zero Temperature Correlators:} 
In Model 2A, there are $N$  Majorana operators $\gamma_\alpha$ coupled to the bulk fields. We have found that $[\gamma_\alpha]=1/4$, same as that in the real SYK model \cite{Kit.KITP.1,Kit.KITP.2,SYKRemarks}, therefore, up to an overall factor, the Majorana Green's function in Model 2A also coincides with that in the SYK model in the long-time limit at $T=0$. We may also identify this equivalence from the self-consistent equations. In the real SYK model, the self-energy $Q(\tau)$ is related to the Green's function $G_\gamma(\tau)$ by $Q(\tau)\propto G_\gamma(\tau)^3$, while we have $Q(\tau)\propto g_{\phi,0}(\tau)G_\gamma(\tau)$ in Model 2A. These are not of the same form but $g_{\phi,0}(\tau)$ happens to be proportional to $G_\gamma(\tau)^2$ in the long-time limit at $T=0$, thus they lead to the same solution. We note that, interestingly, it appears from QMC numerics that the large-$N$ scaling dimension $\gamma_\alpha=1/4$ in Model 2A does not change much as we go down to $N=3$. Recall from the previous subsection, QMC found that in the $N=3$ version of Model 2A, the spin operator\footnote{In Model 2A, the operator $S_{\alpha\beta}\propto \rmi\gamma_\alpha\gamma_\beta$ couples directly to the bulk $\phi$ fields. When $N=3$, $S_\alpha$ is defined by $S_\alpha=\frac{1}{2}\epsilon^{\alpha\beta\gamma}S_{\beta\gamma}$ and is proportional to $\rmi\gamma_0\gamma_\alpha$ in either of the fermion-number parity sectors. } $S_\alpha\propto\rmi\gamma_0\gamma_\alpha$ has scaling dimension 0.20. This quantity does not depend on whether we impose the fermion-number parity projection or not as explained in Section~\ref{SpinorImpTensorBulk}. Noticing that $\gamma_0$ completely decouples from all other degrees of freedom, we conclude $[\gamma_\alpha]=0.20$ when $N=3$, quite closed to its large-$N$ limit $1/4$. 

The equivalence between Model 2B and the complex SYK model \cite{SubirCplxSYK,GuCplxSYK} is even more remarkable. The complex fermion Green's function in Model 2B is given by 
\begin{align}
	\ex{T c_\alpha(\tau) c^\da_\alpha(0)}=\frac{A\sign(\tau)+B}{2|\tau|^{1/2}}\quad (\tau\gg\Lambda_{\rm UV}^{-1},~T=0). 
\end{align}
Apart from the fermion scaling dimension $[c_\alpha]=1/4$ and the overall normalization, this two-point function contains one more parameter $B/A$ characterizing the particle-hole asymmetry, which is determined by the Luttinger theorem introduced in Section~\ref{TensorImpScalingAnsatzSubSec}. Let us quote the result again: 
\begin{align}
	\nu-\frac{1}{2}=-\frac{\theta}{\pi}+\left(\Delta-\frac{1}{2}\right)\frac{\sin(2\theta)}{\sin(2\pi\Delta)}, 
\end{align}
where $\Delta=1/4$ is the fermion scaling dimension and $B/A$ is related to $\theta$ by $B/A=\tan(\theta)/\tan(\pi\Delta)$. It turns out that not only do the fermion scaling dimensions match but the above Luttinger theorem also coincides exactly with the charge formula in the complex SYK model \cite{GeorgesParcolletSachdev,SubirCplxSYK,GuCplxSYK}. This is a rather nontrivial result since the Luttinger theorem presented here contains an anomalous term, the second term on the right-hand side, which comes from certain singularities at $\omega=0$ as detailed in Appendix~\ref{LuttingerThm}. For example, such an anomalous term does not exist in the context of the multichannel $SU(N)$ Kondo problem \cite{SUNKondo}. A deeper understanding of the unexpected similarities between these models awaits future investigation. 

{\bf Finite Temperature:} The equivalence between Model 2A/2B with the real/complex SYK model in the long-time limit does not seem to hold when $T>0$. In the SYK models, the Green's functions at $T>0$ can be obtained from the zero-temperature ones by a simple conformal transformation. However, in our impurity problems, we have an intrinsically 2+1D bulk CFT such that the correlation functions at different temperatures are not simply related by conformal transformations as in 1+1D or 0+1D (nearly) CFTs. 
\section{Conclusion}\label{Conclusions}
In this paper, we investigated four $SO(N)$ symmetric models of a quantum impurity coupled to a 2+1D critical bulk using $1/N$ expansions. Models 1, 2A and 2B are large-$N$ generalizations of $SO(3)$ symmetric models. More specifically, Models 1 and 2A reduce to the problem of a spin-$\frac{1}{2}$ impurity coupled to the $O(3)$ Wilson-Fisher bulk CFT when $N=3$. On the other hand, Model 2B together with $\nu=1/N$ reduces to the problem of a spin-$1$ impurity coupled to the same bulk. Model 3 describes the $O(3)$ Wilson-Fisher bulk coupled to a spin $S>1/2$ impurity  that respects the $\mathbb{Z}_2: \phi_\alpha \to -\phi_\alpha$ symmetry of the bulk. 

For Models 1, 2A and 2B, we found that the impurity is not fully screened by the bulk and exhibits a Curie form static susceptibility with a renormalized Curie coefficient. The absence of screening can be understood using symmetry fractionalization arguments. Model 3 has a rich phase diagram: it possesses a free spin phase for all $S$. In addition, for $S = 1$ it has a fully screened phase separated from the free spin phase by a transition that we describe. For $S \ge 2$ we find a transition from the free spin phase to a yet undetermined phase. 

Let us think about the implications for $N=3$. Extrapolating to $N=3$, our results for Models 1 and 2A both imply that a spin-$\frac{1}{2}$ impurity is not fully screened by the $O(3)$ critical bulk. How about a spin-$1$ impurity? From Model 2B, we have found that in the large-$N$ limit, either a vector impurity ($\nu\rightarrow0$) or an impurity with $\nu=1/3$ has a nonzero $\mc C_{\rm cp}/\mc C_{\rm free}$ and is therefore {\em also} not fully screened by the $O(N(N-1)/2)$ critical bulk generated by the tensor fields $\phi_{\alpha\beta}$. Both cases reduce to $S=1$ when $N=3$. The existence of a stable not fully screened fixed point for $S = 1$ was also confirmed using QMC and  $\epsilon$-expansion, which we discussed in Section~\ref{DifferentApproaches}. In principle, for integer spin $S$ there should also exist a fully screened impurity fixed point, however, it is not accessible in our large-$N$ treatment of Model 2B. In contrast, the transition to what we believe to be a fully screened phase is accessible for $S=1$ in Model 3 and also in its large-$N$ generalizations. 

Let us now comment on other possible future directions. An obvious open problem is to upgrade the impurity to 1+1D, i.e. to couple a 1+1D gapless spin chain to the boundary of a 2+1D critical bulk. Indeed this problem was one of the  motivations for this work. This problem was recently investigated in Refs.~\cite{Jian_gSPT2020, Max3DBCFT}.

 We have  found that Models 2A and 2B are closely related to the real and complex SYK models, respectively. In particular, the Luttinger theorem we proved in Model 2B takes exactly the same form as that in the complex SYK model. It would be  interesting to further explore the connections between our impurity problems and the SYK models. For example, it is presently unclear if the impurity residual entropy in Models 2A and 2B matches with the SYK results. A beautiful formula 
\begin{align}
\frac{\rmd S_{\rm res}}{\rmd\nu}=2\pi\mc E
\end{align}
which relates the entropy $S_{\rm res}$, charge $\nu$ and an ``electric field'' parameter $\mc E$ characterizing particle-hole asymmetry has been proved in both the multichannel $SU(N)$ Kondo problem \cite{SUNKondo} and the complex SYK model \cite{SubirCplxSYK,GuCplxSYK}, although the relation between $\mc E$ and $\nu$ is different in these two cases. We wonder if the same holds in Model 2B where $\mc E=\ln\left[ (1+B/A)/(1-B/A) \right]$; so far we were not able to make progress on this question due to the lack of analytical control of the finite-temperature Green's function. 

\section*{Acknowledgements} 
We would like to thank Ruihua Fan, Subir Sachdev and Matthias Vojta for insightful discussions. HS, SL and AV were supported by a Simons Investigator award (AV) and by the Simons Collaboration on Ultra-Quantum Matter, which is a grant from the Simons Foundation (651440,  AV).  MM is supported by the National Science Foundation under grant number DMR-1847861.

\appendix

\begin{widetext}
\section{Analytic Expressions for the Static Susceptibility}\label{SusceptibilityExpression}
In this section, we explicitly spell out the analytic expressions for the interacting Curie coefficients mentioned in the main text. 

First consider the case of a spinor impurity coupled to vector bulk fields studied in Section~\ref{SpinorImpVecBulk}, recall there is the decomposition: 
\begin{align}
	\chi_{\rm imp}=\frac{\mathcal{C}_{\rm cp}}{T}=\chi_{\rm b,b}+2\chi_{\rm b,imp}+\chi_{\rm imp,imp}. 
\end{align}
We have the following expressions (cf. \figref{SusceptibilityDiagrams_VecBulk}) for the order-$1/N$ contribution to $\chi_{\rm *}$, denoted as $\chi^{(1)}_{\rm *}$. 
\begin{align}
	T\chi^{(1)}_{b,b}&=I_1+I_2+I_3, \\
	I_1+I_2&=-\frac{1}{\pi}\sum_{\omega} B_{\rm reg}(\rmi\omega_n)\frac{\tilde\omega_n^2}{(\tilde\omega_n^2+\mu^2)^2}, \\
	I_3&=\frac{1}{2\pi}\sum_\omega B_{\rm reg}(\rmi\omega_n)\frac{1}{\tilde\omega_n^2+\mu^2}=0, 
\end{align}
where $B_{\rm reg}(\rmi\omega_n)$ represents the shaded bubble and is given by 
\begin{align}
B_{\rm reg}(\rmi\omega_n)&=\frac{1}{N}\frac{\sum_{\lambda,\nu}(\tilde\lambda_n^2+\mu^2)^{-1}\Phi(\rmi\tilde\nu_n)[\Phi_0(\rmi\tilde\omega_n-\rmi\tilde\nu_n)-\Phi_0(\rmi\tilde\lambda_n-\rmi\tilde\nu_n)]}{\sum_\lambda(\tilde\lambda_n^2+\mu^2)^{-1}}. 
\end{align}
In deducing the above result, the following leading-order expression for the $\lambda$ field propagator has been used \cite{ZinnJustinReview}: 
\begin{align}
\Delta_\lambda(\rmi\nu_n,\bs q)&=-\frac{2}{N}\left[ \frac{1}{\beta}\sum_{\omega_n}\int\frac{\rmd^2p}{(2\pi)^2}g_\phi(\rmi\omega_n,\bs p)g_\phi(\rmi\nu_n-\rmi\omega_n,\bs q-\bs p) \right]^{-1}\\
&=
\begin{cases}
-\frac{16}{N}\sqrt{\nu_n^2+q^2} & (m=0), \\
-\frac{16\pi}{\sqrt{5}N}m & (\nu_n=0,~\bs q=0), \\
\cdots &  
\end{cases}
\end{align}
where the minus sign reveals the fact that $\lambda$ has imaginary fluctuation around the saddle point solution. For $\chi_{\rm b,imp}$ we have
\begin{align}
	2T\chi^{(1)}_{\rm b,imp}&=\frac{2}{\pi N}\sum_{\omega,\nu}\frac{\tilde\omega_n+\tilde\nu_n}{(\tilde\omega_n+\tilde\nu_n)^2+\mu^2}\frac{1}{\tilde\omega_n^2}\imag \Phi_0(\rmi\tilde \nu_n). 
\end{align}
For $\chi_{\rm imp,imp}$, we have
\begin{align}
	T\chi^{(1)}_{\rm imp,imp}=\frac{2}{\pi N}\sum_{\omega>0}\sum_{\nu>0}\frac{1}{\tilde\omega_n^3}\imag\Phi_0(\rmi\tilde\nu_n)\ln\left[ \frac{(\tilde\omega_n-\tilde\nu_n)^2+\mu^2}{(\tilde\omega_n+\tilde\nu_n)^2+\mu^2} \right]. 
\end{align}
As we mentioned in the main text, there is a divergence cancellation between $\chi^{(1)}_{\rm imp,imp}$ and $2\chi^{(1)}_{\rm b,imp}$, which is not obvious from the analytic expressions above. 

For the case of a spinor impurity coupled to tensor bulk fields studied in Section~\ref{SpinorImpTensorBulk}, the analytic expression for $\mc C_{\rm cp}$ is (cf. \figref{SusceptibilityDiagrams}) 
\begin{align}
\mc C_{\rm cp}&=I_1+I_2+I_3+I_4, \\
I_1+I_2&=-\frac{1}{\pi}\sum_{\omega}[N B_{\rm reg}(\rmi\omega_n)]\frac{\tilde\omega_n^2}{(\tilde\omega_n^2+\mu^2)^2}, \\
I_3&=\frac{1}{4\pi^2}\sum_{\omega,\lambda,\nu}\frac{(\tilde\nu_n-\tilde\omega_n)(\tilde\nu_n-\tilde\lambda_n)}{[(\tilde\nu_n-\tilde\omega_n)^2+\mu^2][(\tilde\nu_n-\tilde\lambda_n)^2+\mu^2]}\Phi^2_\gamma(\rmi\tilde\nu_n)\Phi_\gamma(\rmi\tilde\omega_n)\Phi_\gamma(\rmi\tilde\lambda_n), \\
I_4&=\frac{1}{2\pi}\sum_\omega[N B_{\rm reg}(\rmi\omega_n)]\frac{1}{\tilde\omega_n^2+\mu^2}=0, 
\end{align}
where $B_{\rm reg}(\rmi\omega_n)$ represents the shaded bubble and is given by 
\begin{align}
N B_{\rm reg}(\rmi\omega_n)&=\frac{\sum_{\lambda,\nu}(\tilde\lambda_n^2+\mu^2)^{-1}\Phi_\gamma(\rmi\tilde\nu_n)[\Phi_\gamma(\rmi\tilde\omega_n-\rmi\tilde\nu_n)-\Phi_\gamma(\rmi\tilde\lambda_n-\rmi\tilde\nu_n)]}{\sum_\lambda(\tilde\lambda_n^2+\mu^2)^{-1}}. 
\end{align}

For the case of an antisymmetric $\nu N$-tensor impurity coupled to tensor bulk fields, we have
\begin{align}
\mc C_{\rm cp}&=I_1+I_2+I_3+I_4, \\
I_1+I_2&=-\frac{1}{\pi}\sum_{\omega}[N B_{\rm reg}(\rmi\omega_n)]\frac{\tilde\omega_n^2}{(\tilde\omega_n^2+\mu^2)^2}, \\
I_3&=\frac{1}{2\pi^2}\sum_{\omega,\lambda,\nu}\frac{(\tilde\nu_n-\tilde\omega_n)(\tilde\nu_n-\tilde\lambda_n)}{[(\tilde\nu_n-\tilde\omega_n)^2+\mu^2][(\tilde\nu_n-\tilde\lambda_n)^2+\mu^2]}\real\left[F^2(\rmi\tilde\nu_n)F(\rmi\tilde\omega_n)F(\rmi\tilde\lambda_n)\right], \\
I_4&=\frac{1}{2\pi}\sum_\omega[N B_{\rm reg}(\rmi\omega_n)]\frac{1}{\tilde\omega_n^2+\mu^2}=0, 
\end{align}
where
\begin{align}
N B_{\rm reg}(\rmi\omega_n)&=\frac{2\sum_{\lambda,\nu}(\tilde\lambda_n^2+\mu^2)^{-1}\real\{F(\rmi\tilde\nu_n)[F(\rmi\tilde\nu_n-\rmi\tilde\omega_n)-F(\rmi\tilde\nu_n-\rmi\tilde\lambda_n)]\}}{\sum_\lambda(\tilde\lambda_n^2+\mu^2)^{-1}}. 
\end{align}

\section{A Luttinger Theorem for the Impurity Problem}\label{LuttingerThm}
Here we derive a Luttinger-like theorem which relates the fermion number constraint \eqref{FermionNumEq} at UV to the IR behavior of the Green's functions. The derivation here is inspired by \cite{SUNKondo}. 

We first introduce a convenient alternative representation of the Green's function and the self-energy function. Note that $G_A(\tau)$ being real antisymmetric, $G_{AB}(\tau)$ being imaginary symmetric and $g_{\phi,0}(\tau)$ being real symmetric altogether imply that both $G(\rmi\omega_n)$ and $Q(\rmi\omega_n)$ are purely imaginary. Moreover, $\imag G(\rmi\omega_n)$ and $\imag Q(\rmi\omega_n)$ both have the form $a\mathbbm{1}+b(\rmi\sigma_y)$ for $a,b\in\mathbb{R}$. There is an isomorphism of (algebraic) fields: 
\begin{align}
\{ a\mathbbm{1}+b(\rmi\sigma_y)|a,b\in\mathbb{R} \}&\cong \mathbb{C}. \\
a\mathbbm{1}+b(\rmi\sigma_y)&\mapsto a+b\rmi\nonumber
\end{align}
Therefore, we can define complex valued functions 
\begin{align}
D(\rmi\omega_n)&=D_1+\rmi D_2=\imag G_A(\rmi\omega_n)+\rmi \imag G_{AB}(\rmi\omega_n), \\
\Sigma(\rmi\omega_n)&=\Sigma_1+\rmi \Sigma_2=\imag Q_A(\rmi\omega_n)+\rmi \imag Q_{AB}(\rmi\omega_n), 
\end{align}
and they satisfy the same algebra as the two-by-two matrices $\imag G(\rmi\omega_n)$ and $\imag Q(\rmi\omega_n)$ (but one need to be careful when taking traces). In particular, the self-consistent equations for them have the same form. 

Let us now derive the Luttinger-like theorem. We consider $T=0$, and the fermion number constraint can be written as 
\begin{align}
-\int\frac{\rmd\omega}{2\pi}\imag D(\rmi\omega)=2\nu-1. 
\label{FermionNumEq_D_T0}
\end{align}
We do not know the full functional form of $D(\rmi\omega)$, so our goal is to evaluate the left-hand side only using the behavior of $D(\rmi\omega)$ at $\omega\rightarrow0\pm$ and $\omega\rightarrow\pm\infty$. 
The self-consistent equations can be written as 
\begin{align}
D^{-1}(\rmi\omega)&=\frac{1}{2}\omega-\rmi\xi+\Sigma(\rmi\omega), \label{SelfConsEq1_DSigma_T0}\\
\Sigma(\rmi\omega)&=J_0^2\int\frac{\rmd\nu}{2\pi}g_{\phi,0}(\rmi\nu)D(\rmi\omega-\rmi\nu).\label{SelfConsEq2_DSigma_T0}
\end{align}
Taking $\omega$ derivative on the first equation, we found 
\begin{align}
D=-2\left( D^{-1}\frac{\pa D}{\pa \omega}+D\frac{\pa\Sigma}{\pa\omega} \right). 
\end{align}
\eqnref{FermionNumEq_D_T0} can now be written as 
\begin{align}
2\int\frac{\rmd\omega}{2\pi}\imag\left( D^{-1}\frac{\pa D}{\pa \omega}+D\frac{\pa\Sigma}{\pa\omega} \right)=2\nu-1. 
\end{align}
We separate the left-hand side into two integrals. In doing so, we actually need to specify how to deal with the singularities of both integrands at $\omega=0$. We choose\footnote{One can also regularize both $D(\rmi\omega)$ and $\Sigma(\rmi\omega)$ near $\omega=0$ in a way compatible with \eqref{SelfConsEq1_DSigma_T0} and that both integrands become smooth. As a result, values of the two integrations will be different from our choice, although the sum of them is still be same. } to break $\int_{-\infty}^{\infty}\rmd\omega\mapsto\int_{-\infty}^{0-}\rmd\omega+\int_{0+}^{\infty}\rmd\omega$. The first integral is straightforward: 
\begin{align}
\iota_1&=2\left( \int_{-\infty}^{0-}+\int_{0+}^\infty \right)\frac{\rmd\omega}{2\pi}\imag \left(D^{-1}\frac{\pa D}{\pa \omega}\right)=\frac{1}{\pi}\left( \int_{-\infty}^{0-}+\int_{0+}^\infty \right)\rmd\omega\frac{\pa}{\pa\omega}\imag\ln D\nonumber\\
&=\frac{1}{\pi}(\theta_{\infty}-\theta_{0+}+\theta_{0-}-\theta_{-\infty}), 
\end{align}
where $\theta_\omega$ is the phase angle of $D(\rmi\omega)$. By the symmetries of the Green's functions, $D(-\rmi\omega)=-D(\rmi\omega)^*$, thus we can take $\theta_{-\omega}=\pi-\theta_{\omega}$ and the above simplifies to $\iota_1=(2/\pi)(\theta_\infty-\theta_{0+})$. \eqref{SelfConsEq1_DSigma_T0} implies that as $|\omega|\rightarrow\infty$, $D(\rmi\omega)\rightarrow 2/\omega$. We thus have $\iota_1=(-2/\pi)\theta_{0+}$ by taking $\theta_{\infty}=0$. At small frequency, we have
\begin{align}
D(\rmi\omega)\rightarrow 2[A\cos(\pi\Delta)\sign(\omega)+\rmi B\sin(\pi\Delta)]\Gamma(1-2\Delta)|\omega|^{2\Delta-1},  
\end{align}
hence, 
\begin{align}
(\cos\theta_{0+},\sin\theta_{0+})=\frac{1}{\sqrt{A^2\cos^2(\pi\Delta)+B^2\sin^2(\pi\Delta)}}(A\cos(\pi\Delta),B\sin(\pi\Delta)). 
\end{align}
From our previous result, $A^2\cos^2(\pi\Delta)+B^2\sin^2(\pi\Delta)=\delta\sin(\pi\delta)/(4\pi J_0^2C_\phi)$ is a constant independent of $\nu$. Also recall that we assume $1/2\leq\delta<1$, implying $0<\Delta\leq1/4$. 

Next, let us analyze the second integral
\begin{align}
\iota_2=2\left( \int_{-\infty}^{0-}+\int_{0+}^\infty \right)\frac{\rmd\omega}{2\pi}\imag\left(D\frac{\pa\Sigma}{\pa\omega} \right), 
\end{align}
which turns out to be tricky. We will first describe an na\"{i}ve approach leading to a wrong result, and then explain how to fix it. Define a Luttinger-Ward functional
\begin{align}
\Phi_{\rm LW}=\int\rmd\tau g_{\phi,0}(\tau)\Tr[\sigma_yG(-\tau)G(\tau)]=-2\rmi\int\frac{\rmd\nu\rmd\omega}{(2\pi)^2}g_{\phi,0}(\rmi\nu)\imag[D(\rmi\omega)D(\rmi\omega-\rmi\nu)]. 
\end{align} 
$\Phi_{\rm LW}$ is a finite number; it is in fact zero due to the symmetries of the Green's functions. $\Phi_{\rm LW}$ should be invariant by $\omega\mapsto\omega+\epsilon$, which implies
\begin{align}
\int\frac{\rmd\nu\rmd\omega}{(2\pi)^2}g_{\phi,0}(\rmi\nu)\imag\left[ \frac{\pa D(\rmi\omega)}{\pa\omega}D(\rmi\omega-\rmi\nu)+D(\rmi\omega)\frac{\pa D(\rmi\omega-\rmi\nu)}{\pa\omega} \right]=0. 
\end{align}
Applying the change of variable $\omega=\omega'+\nu$ to the second term in the square bracket, and noticing \eqnref{SelfConsEq2_DSigma_T0}, the above reduces to 
\begin{align}
\int\frac{\rmd\omega}{2\pi}\imag\left[ \frac{\pa D(\rmi\omega)}{\pa\omega}\Sigma(\rmi\omega) \right]=0\quad\Rightarrow\quad
\int\frac{\rmd\omega}{2\pi}\imag\left[D(\rmi\omega) \frac{\pa \Sigma(\rmi\omega)}{\pa\omega} \right]=0, 
\end{align}
where integration by parts was used in the last step. Does the above calculation imply $\iota_2=0$? Suppose this is true, we conclude that $\theta_{0+}=(1/2-\nu)\pi$. As a result, $A=0$ for $\nu=0$ and the impurity spin correlation function (at long time) $G_S(\tau)=\frac{1}{2}(A^2-B^2)/|\tau|^{4\Delta}$ becomes negative in this case, which is not possible given that $S_{\alpha\beta}$ is a bosonic hermitian operator. Therefore, the integration in the above result, whatever it means, is not equivalent to that in $\iota_2$. In fact, for this calculation to be valid, we need to regularize the singularity of $D(\rmi\omega)$ near $\omega=0$, which also modifies $\Sigma(\rmi\omega)$ through \eqref{SelfConsEq2_DSigma_T0}. The regularization leaves a contribution and makes $\iota_2$ nonzero. Notice that although both $G_A\pa R/\pa\omega$ and $G_{AB}\pa Q_A/\pa\omega$ contain $\mc O(|\omega|^{-1})$ terms, these terms cancel in $\imag(D\pa \Sigma/\pa \omega)$ and the integral in $\iota_2$ has no logarithmic divergence near $\omega=0$. Similar cancellation may no longer exist for the regularization; suppose we regularize $D(\rmi\omega)$ within $|\omega|\lesssim \epsilon$, then there is a contribution of order $\epsilon\times(1/\epsilon)\sim 1$ to $\int\rmd\omega\imag(D\pa\Sigma/\pa\omega)$, hence $\iota_2$ should be of order $1$. According to this argument, in order to get rid of the effect of regularization, we need to somehow reduce the small-frequency divergence in $D$ or $\pa\Sigma/\pa\omega$. We can achieve this goal by utilizing the $D$ and $\Sigma$ functions at the special filling $\nu=0$, denoted by $D_0$ and $\Sigma_0$. Let $A_0,B_0$ be the $A,B$ coefficients at $\nu=0$. For a general filling, we define
\begin{align}
\tilde D&=\frac{A}{A_0}\real D_0+\rmi\frac{B}{B_0}\imag D_0, \\
\tilde \Sigma&=\frac{A}{A_0}\real \Sigma_0+\rmi\frac{B}{B_0}\imag \Sigma_0, 
\end{align}
Then $D-\tilde D$ as well as $\pa(\Sigma-\tilde\Sigma)/\pa\omega$ no longer have $\mc O(|\omega|^{-1/2})$ terms. Now we can perform a similar calculation. Define a new Luttinger-Ward functional: 
\begin{align}
\Phi'_{\rm LW}=\int\frac{\rmd\nu\rmd\omega}{(2\pi)^2}g_{\phi,0}(\rmi\nu)\imag[(D+\tilde D)(\rmi\omega)(D-\tilde D)(\rmi\omega-\rmi\nu)]. 
\end{align}
By the invariance under $\omega\mapsto\omega+\epsilon$, we obtain the sum rule
\begin{align}
\int\frac{\rmd\omega}{2\pi}\imag\left[ (D-\tilde D)\frac{\pa(\Sigma+\tilde\Sigma)}{\pa\omega}+(D+\tilde D)\frac{\pa(\Sigma-\tilde\Sigma)}{\pa\omega} \right]=0.   
\end{align}
Here we can substitute $\int_{-\infty}^{\infty}\rmd\omega$ by $\int_{-\infty}^{0-}\rmd\omega+\int_{0+}^{\infty}\rmd\omega$, and the above simplifies to 
\begin{align}
\left( \int_{-\infty}^{0-}+\int_{0+}^{\infty} \right)\frac{\rmd\omega}{2\pi}\imag\left[ D\frac{\pa\Sigma}{\pa\omega}-\tilde D\frac{\pa\tilde\Sigma}{\pa\omega} \right]=0. 
\end{align}
Noticing
\begin{align}
\imag\left( \tilde D\frac{\pa\tilde \Sigma}{\pa\omega} \right)=\frac{AB}{A_0B_0}\imag\left( D_0\frac{\pa\Sigma_0}{\pa\omega} \right), 
\end{align}
we have
\begin{align}
\iota_2(\nu)=\frac{AB}{A_0B_0}\iota_2(\nu=0). 
\end{align}
% We can compute $\iota_2$ for a general $\nu$ as long as we figure out its value at $\nu=0$. 
For $\nu=0$, the impurity Hilbert space is trivial, thus $G_S(\tau)$ should be identically zero, implying $A_0^2=B_0^2\neq 0$. We know that $A_0>0$, but the sign of $B_0$ can be arbitrary. Let us assume $B_0>0$, which may follow from the assumption that $\imag G_{AB}(\tau)$ does not change sign at zero temperature, or can be checked numerically, then $\theta_{0+}(\nu=0)=\pi\Delta+2n\pi$ for some integer $n$. Using $\iota_1+\iota_2=2\nu-1$, we obtain $\iota_2(\nu=0)=4n+2\Delta-1$. Combining all these results, it follows that
\begin{align}
-\frac{2}{\pi}\theta_{0+}+\frac{\sin(2\theta_{0+})}{\sin(2\pi\Delta)}(4n+2\Delta-1)=2\nu-1, 
\end{align}
where $AB/(A_0B_0)=\sin(2\theta_{0+})/\sin(2\pi\Delta)$ has been used. The system has a particle-hole symmetry: $\gamma_A\mapsto\gamma_A,\gamma_B\mapsto-\gamma_B,\xi\mapsto-\xi$ and $\nu\mapsto 1-\nu$, which implies $A(\nu)=A(1-\nu)$ and $B(\nu)=-B(1-\nu)$. Assuming the continuity of $\theta_{0+}$ as a function of $\nu$, this symmetry and $A$ being nonnegative imply that $-\pi\Delta+2n\pi\leq \theta_{0+}\leq \pi\Delta+2n\pi$ as $\nu$ varies. In particular, $\theta_{0+}(\nu=1/2)=2n\pi$. Plugging this into the above equation yields $n=0$. We finally conclude that $\theta_{0+}$ satisfies the following transcendental equation: 
\begin{align}
\frac{\sin(2\theta_{0+})}{\sin(2\pi\Delta)}=\frac{1}{2\Delta-1}\left( 2\nu-1+\frac{2}{\pi}\theta_{0+} \right), 
\end{align}
which then determines the ratio $B/A=\tan(\theta_{0+})/\tan(\pi\Delta)$.

\section{Perturbative calculations for the $\mathbb{Z}_2$ symmetric impurity}
\label{app:PertTensor}
We start with the $O(N)$ symmetric model in Eq.~(\ref{OtN}). We normalize the $O(N)$ traceless symmetric operator $t_{\alpha_1 \alpha_2}$ as

\begin{align}
\langle t_{\alpha_1 \alpha_2}(x_1) t_{\beta_1 \beta_2}(x_2)\rangle &= \frac{1}{x^{2 \Delta_t}_{12}} \frac{1}{2} \left(\delta_{\alpha_1 \beta_1} \delta_{\alpha_2 \beta_2} + \delta_{\alpha_1 \beta_2} \delta_{\alpha_2 \beta_1} -\frac{2}{N} \delta_{\alpha_1 \alpha_2} \delta_{\beta_1 \beta_2}\right) \label{tnorm} \\
t_{\alpha_1 \alpha_2}(x_1) t_{\beta_1 \beta_2}(x_2) &\sim \frac{1}{x_{12}^{\Delta_t}} \frac{\lambda_{ttt}}{2 \sqrt{2}} \left(\delta_{\alpha_1 \beta_1} t_{\alpha_2 \beta_2} + \delta_{\alpha_1 \beta_2} t_{\alpha_2 \beta_1} + \delta_{\alpha_2 \beta_1} t_{\alpha_1 \beta_2} + \delta_{\alpha_2 \beta_2} t_{\alpha_1 \beta_1} - \frac{4}{N} \delta_{\alpha_1 \alpha_2} t_{\beta_1 \beta_2} - \frac{4}{N} \delta_{\beta_1 \beta_2} t_{\alpha_1 \alpha_2}\right) \label{TOPE}
\end{align}
Our normalization here agrees with Ref.~\onlinecite{O3bootstrap}. Note, we have only included the operator $t$ on the RHS of the OPE (\ref{TOPE}). Also, we did not include any descendants.

Now, starting from Eq.~(\ref{OtN}) and performing perturbative RG in $J$, 

\begin{align} \delta S &= -\frac{J^2}{2} \int d\tau_1 d \tau_2 {\cal T}\{O_{\alpha_1 \alpha_2}(\tau_1) O_{\beta_1 \beta_2}(\tau_2) \} t_{\alpha_1 \alpha_2}(\tau_1) t_{\beta_1 \beta_2}(\tau_2) 
\to -\frac{J^2 \lambda_{ttt}}{\sqrt{2}}\int \frac{d\tau_1 d \tau_2}{|\tau_1 - \tau_2|^{\Delta_t}} {\cal T} \{O_{\gamma \alpha} (\tau_1) O_{\gamma \beta}(\tau_2) \} t_{\alpha \beta}\left(\frac{\tau_1 + \tau_2}{2}\right) 
\label{SRG}
\end{align}
Here ${\cal T}$ stands for time-ordering and we used the OPE (\ref{TOPE}) in the last step. Now 
\begin{align} O_{\gamma \alpha}O_{\gamma \beta} + O_{\gamma \beta} O_{\gamma \alpha} = c(N, N_b) O_{\alpha \beta} + d(N, N_b) \delta_{\alpha \beta}\end{align}
where
\begin{align} c(N,N_b) &= - \frac{1}{2N} (4 (N-4) N^2_b + 4(N^2 -6 N +8) N_b + N (N^2 -4N +8)) \nonumber \\
d(N, N_b) &= \frac{1}{N^2} (N-2) N_b (N_b + N - 2) (2 N_b + N-4) (2N_b+N) \label{cd} \end{align}
This result can be obtained by using the representation of the impurity in terms of $b_\alpha$, $b^{\dagger}_\alpha$ operators discussed in section \ref{TensorSymmSaddle}. For the $O(3)$ case $N_b = S$ - the spin of the impurity and Eq.~(\ref{cd}) also works for half-integer $S$.
Now integrating over small $\tau_1 - \tau_2$ in (\ref{SRG}) and using $\Delta_t \approx 1$, we obtain
\begin{align} \frac{d J}{d \ell} = \frac{ \lambda_{ttt} c(N,N_b)}{\sqrt{2}} J^2\end{align}

We next compute the anomalous dimension of an operator $A$ acting on the impurity. We will be mostly interested in the $N = 3$ case with $A = S_\alpha$. Consider a correlation function with $A$ insertion, $\langle A(\tau) \ldots\rangle$, where ellipses stand for other operators. Performing perturbation theory in $J$:
\begin{align} \langle A(\tau) \ldots\rangle - \langle A(\tau) \ldots \rangle_0 &=  + \frac{J^2}{2} \int d\tau_1 d \tau_2 \big(\langle {\cal T}\{A(\tau) O_{\alpha_1 \alpha_2}(\tau_1) O_{\beta_1 \beta_2}(\tau_2)\ldots \}\rangle_0 -  \langle {\cal T}\{A(\tau) \ldots \}\rangle_0 \langle {\cal T} \{ O_{\alpha_1 \alpha_2}(\tau_1) O_{\beta_1 \beta_2}(\tau_2) \}\rangle_0 \big) 
\nonumber\\ &\times \langle t_{\alpha_1 \alpha_2}(\tau_1) t_{\beta_1 \beta_2} (\tau_2)\rangle_0 \end{align} 
The expectation values on the RHS are understood to be taken in the $J = 0$ theory. From now on, we omit ellipses. Using (\ref{tnorm}),
\begin{align} 
\langle A(\tau) \rangle -  \langle A(\tau) \rangle_0 =  \frac{J^2}{2} \int \frac{d\tau_1 d \tau_2}{|\tau_1 -\tau_2|^{2 \Delta_t}}  \big(\langle {\cal T}\{A(\tau) O_{\alpha \beta}(\tau_1) O_{\alpha \beta}(\tau_2)\}\rangle_0 -  \langle {\cal T}\{A(\tau)\}\rangle_0 \langle {\cal T} \{ O_{\alpha \beta}(\tau_1) O_{\alpha \beta}(\tau_2) \}\rangle_0 \big) 
\end{align}
Now, $O_{\alpha \beta} O_{\alpha \beta} = \frac12 N d(N, N_b)$. Thus, the RHS is only non-vanishing when $\tau_1$ and $\tau_2$ are on different sides of $\tau$:
\begin{align} 
\langle A(\tau) \rangle- \langle A(\tau) \rangle_0 = J^2 \int_{\tau}^{\infty} d\tau_1 \int_{-\infty}^{\tau} d \tau_2 \frac{1}{|\tau_1 -\tau_2|^{2 \Delta_t}}  \langle [O_{\alpha \beta}, A] O_{\alpha \beta}\rangle_0 \sim J^2 \log \Lambda \langle [O_{\alpha \beta}, A] O_{\alpha \beta}\rangle_0 
\end{align}
where we've isolated the $UV$ divergent part coming from $\tau_1$, $\tau_2$ close to $\tau$.
Now, let's take $A = S_{\gamma \delta}$. We have, 
\begin{align} [O_{\alpha \beta}, S_{\gamma \delta}] O_{\alpha \beta} = 2 i [O_{\alpha \gamma}, O_{\alpha \delta}] = - \frac{1}{2} (N-2) (2 N_b + N) (2N_b + N - 4) S_{\gamma \delta} \end{align}
which can again be obtained using the representation of $O_{\alpha \beta}$ in terms of $b$, $b^{\dagger}$.
Thus, the anomalous dimension
\begin{align} \Delta_{S_{\alpha \beta}} = \frac{J^2}{2}(N-2) (2 N_b + N) (2N_b + N - 4) \end{align}
which coincides with (\ref{DeltaST}) for $N = 3$.
\end{widetext}

\bibliography{Bib_LargeN.bib,Bib_LargeNAV.bib}
\end{document}